\pdfoutput=1

\documentclass[11pt,nofootinbib]{article}
\usepackage{amsmath,amssymb,mathtools}
\usepackage{graphicx}
\usepackage[colorlinks=true,linkcolor=blue,citecolor=blue,urlcolor=blue,linktocpage=true]{hyperref}
\usepackage{subfigure}
\usepackage{comment}
\usepackage{tensor}
\usepackage{cases}
\usepackage[utf8]{inputenc}

\numberwithin{equation}{section}

\newcommand{\beq}{\begin{equation}}
\newcommand{\eeq}{\end{equation}}
\newcommand{\beqa}{\begin{eqnarray}}
\newcommand{\eeqa}{\end{eqnarray}}
\newcommand{\bea}{\begin{eqnarray}}
\newcommand{\eea}{\end{eqnarray}}

\begin{document}

\providecommand{\abs}[1]{\lvert#1\rvert}
\providecommand{\bd}[1]{\boldsymbol{#1}}

\begin{titlepage}

\setcounter{page}{1} \baselineskip=15.5pt \thispagestyle{empty}

\begin{flushright}
\end{flushright}
\vfil

\bigskip
\begin{center}
 {\LARGE \textbf{Early Cosmological Evolution of}}\\
 \medskip
 {\LARGE \textbf{Primordial Electromagnetic Fields}} 
\vskip 15pt
\end{center}

\vspace{0.5cm}
\begin{center}
{\Large
Takeshi Kobayashi$^{\star}$
and
Martin S. Sloth$^{\dagger}$
}\end{center}

\vspace{0.3cm}

\begin{center}
\textit{$^{\star}$ International Centre for Theoretical Physics,\\
 Strada Costiera 11, 34151 Trieste, Italy}\\

\vskip 14pt 
\textit{$^{\dagger}$ CP$^3$-Origins, Center for Cosmology and Particle
 Physics Phenomenology,\\
 University of Southern Denmark, Campusvej 55, 5230 Odense M, Denmark}\\ 
 
\vskip 14pt
E-mail:
 \texttt{\href{mailto:takeshi@ictp.it}{takeshi@ictp.it}},
 \texttt{\href{mailto:sloth@cp3.sdu.dk}{sloth@cp3.sdu.dk}}
\end{center} 



\vspace{1cm}

\noindent
It is usually assumed that when Weyl invariance is unbroken in the electromagnetic sector, the energy density of primordial magnetic fields will redshift as radiation. Here we show that primordial magnetic fields do {\it not} exhibit radiation-like redshifting in the presence of stronger electric fields, as a consequence of Faraday's law of induction. In particular for the standard Maxwell theory, magnetic fields on super-horizon scales can redshift as $B^2 \propto a^{-6} H^{-2}$, instead of the usually assumed $a^{-4}$. Taking into account this effect for inflationary magnetogenesis can correct previous estimates of the magnetic field strength by up to 37 orders of magnitude. This opens new possibilities for inflationary magnetogenesis, and as an example we propose a scenario where femto-Gauss intergalactic magnetic fields are created on Mpc scales, with high-scale inflation producing observable primordial gravitational waves, and reheating happening at low temperatures.
\vfil

\end{titlepage}

\newpage
\tableofcontents

\section{Introduction}
\label{sec:intro}

The origin of the magnetic fields in our universe is a mystery.
There are several known astrophysical and cosmological mechanisms for
producing the galactic magnetic fields.
On the other hand for intergalactic magnetic fields which are
suggested by recent gamma ray observations to be of femto-Gauss
strength,
their large correlation length (typically of megaparsec scales or
larger) indicates a cosmological
origin~\cite{Tavecchio:2010mk,Neronov:1900zz,Chen:2014rsa}.
Theories of primordial magnetic field generation have been widely studied,
and the proposed mechanisms include magnetogenesis 
during the inflationary epoch~\cite{Turner:1987bw,Ratra:1991bn},
the post-inflationary epoch~\cite{Kobayashi:2014sga},
and upon cosmological phase
transitions~\cite{Vachaspati:1991nm,Cornwall:1997ms}.
See also reviews such as
\cite{Widrow:2002ud,Barrow:2006ch,Durrer:2013pga,Subramanian:2015lua}
and references therein. These proposed mechanisms are however not without challenges. For example, in order for the description to stay perturbative and avoid backreaction \cite{Demozzi:2009fu}, working models of inflationary magnetogenesis leading to femto-Gauss magnetic fields on the megaparsec scale have so far required a very low scale of inflation \cite{Fujita:2012rb,Ferreira:2013sqa,Ferreira:2014hma,Green:2015fss} or a combination of mechanisms \cite{Kobayashi:2014sga}.

Here, in order to connect the magnetic fields produced in the early
universe with those (indirectly) observed in the present universe, it is
crucial to understand the evolution of magnetic fields along the
cosmological history.
In most of the literature on cosmological magnetogenesis, 
it is assumed that magnetic fields on
super-horizon scales
undergo a radiation-like redshifting with the cosmological scale
factor~$a$ as
\begin{equation}
 B^2 \propto \frac{1}{a^4}.
\label{B2a4}
\end{equation}
This rather rapid decay has been considered as the main obstacle against
magnetic fields produced in the primordial universe from surviving until
today and seeding the observed fields. 

However, it is actually the sum of the
magnetic and electric fields $B^2 + E^2$ which redshifts as radiation,
whereas the individual $B^2$ and $E^2$ can have different redshift behaviors;
the goal of our paper is to explicitly show this.
In particular when the electric field is stronger than the magnetic field, 
we show that the magnetic field outside the horizon can evolve in time as
\begin{equation}
 B^2 \propto \frac{1}{a^6 H^2},
\label{B2a6H2}
\end{equation}
where $H$ is the Hubble rate.
In a decelerating universe, this yields less redshift to the magnetic
fields compared to~(\ref{B2a4}).
For instance, if the universe is effectively matter-dominated,
i.e. $H^2 \propto a^{-3}$, 
the magnetic field would redshift as $B^2 \propto a^{-3}$. 
Such a behavior of cosmological magnetic fields was seen
in~\cite{Kobayashi:2014sga} in the context of post-inflationary
magnetogenesis scenarios.\footnote{Most of the analyses in
\cite{Kobayashi:2014sga} are based on 
directly solving the gauge field's equation of motion, arriving at
the correct scaling behavior of the magnetic field.
However their Section~3.2 assumes the redshifting (\ref{B2a4}) and thus
can be modified by taking into account the proper magnetic
scaling.}
In this paper, we show that super-horizon magnetic fields
in a decelerating universe generically follow the scaling~(\ref{B2a6H2})
in the presence of stronger electric fields.\footnote{Non-radiation-like
redshifting of magnetic fields has also been claimed for
anisotropic~\cite{Barrow:1997sy} or open~\cite{Barrow:2008jp} universes,
although the mechanism for the open universe was strongly questioned
in~\cite{Adamek:2011hi}. 
Other proposals exist as well, e.g.~\cite{Tsagas:2016fax}.
However we stress that the effect discussed in the current paper is
different from those.}

Many of the previously proposed inflationary magnetogenesis scenarios,
including the well-studied $I^2 FF$ model~\cite{Ratra:1991bn},
produce much stronger primordial electric fields than magnetic fields
during the inflationary epoch.
The electric fields continue to exist after inflation until the universe turns into a good conductor. This can happen any time from the end of inflation until the end of reheating depending on the details of the reheating mechanism \cite{Turner:1987bw}. It is usually assumed that conductivity turns on already at the end of inflation erasing the electric field, but if the conductivity remains small
during this epoch between the end of inflation and the end of reheating,
the strong electric field induces the magnetic field evolution of~(\ref{B2a6H2}),
which yields less redshift compared to the usually assumed~(\ref{B2a4}).
As a consequence, the present-day amplitude of magnetic fields
arising from inflationary magnetogenesis can actually be much larger than
what has been claimed in previous studies.
The difference is drastic especially when there is a hierarchy
between the inflation and reheating scales;
this implies that a higher inflation scale can help produce
stronger magnetic fields today,
as opposed to the widespread belief based on~(\ref{B2a4}) 
that high-scale inflation is incompatible with efficient inflationary
magnetogenesis. While the conclusions of \cite{Ferreira:2013sqa,Ferreira:2014hma}, that femto-Gauss magnetic fields on the Mpc scale require inflation to happen below the TeV scale\footnote{One can also consider other options, like the generation of helical magnetic fields from a coupling of the type $I^2 F_{\mu \nu}\tilde F^{\mu\nu}$, where $\tilde F$ is the dual field strength \cite{Caprini:2014mja}. Such mechanisms however suffer from their own backreaction, anisotropy and perturbativity constraints \cite{Ferreira:2014zia,Ferreira:2015omg} yielding similar problems for magnetogenesis \cite{Caprini:2017vnn}.}, remains true under their assumption of instantaneous reheating or high conductivity throughout reheating, a prolonged period of reheating with vanishing conductivity can significantly alter these conclusions -- opening a new space for inflationary magnetogenesis phenomenology and model building. 
As an example, we propose a toy model of inflationary magnetogenesis 
that can produce the femto-Gauss intergalactic magnetic fields
during high-scale inflation, while being free from strong
couplings~\cite{Demozzi:2009fu,Gasperini:1995dh} or
affecting the background
cosmology~\cite{Fujita:2012rb,Green:2015fss,Bamba:2003av,Kanno:2009ei} and 
curvature
perturbations~\cite{Ferreira:2013sqa,Ferreira:2014hma,Ferreira:2015omg,Barnaby:2012tk,Suyama:2012wh,Jain:2012ga,Jain:2012vm,Nurmi:2013gpa,Fujita:2014sna,Ganc:2014wia}. 
We demonstrate how the various constraints on primordial 
magnetogenesis claimed in the literature such as those cited here
are relaxed when the electric field-induced scaling~(\ref{B2a6H2}) is
taken into account. 

We also study the cosmological consequences of primordial electric fields.
By analyzing their gravitational backreaction,
we derive constraints on magnetic fields produced from generic Weyl
symmetry-breaking scenarios during the inflationary and
post-inflationary epochs.
Primordial electric fields can also raise the conductivity of the
universe even before reheating by producing charged particles via the
Schwinger process~\cite{Kobayashi:2014zza};
this issue will also be discussed. 

The paper is organized as follows:
In Section~\ref{sec:Faraday} we provide a simple argument for the
electromagnetic scalings based on Faraday's law of induction,
without specifying the gauge field action.
In Section~\ref{sec:Bogoliubov}
we focus on $I^2 FF$ theories and give a more rigorous derivation using
Bogoliubov coefficients.
We study a toy model of inflationary magnetogenesis in
Section~\ref{sec:power-law}, where we see how the induction effect
impacts the final magnetic field strength; here we also propose a
scenario capable of producing femto-Gauss intergalactic magnetic fields
during high-scale inflation.
We further provide model-independent constraints on primordial magnetic
fields in Section~\ref{sec:MIC},
and then briefly discuss the possibility of electric field quenching due
to the Schwinger process in Section~\ref{sec:Schwinger}.
We summarize our findings in Section~\ref{sec:conc}.

We will occasionally use the conversions of
$1\, \mathrm{G} \approx 2 \times 10^{-20} \, \mathrm{GeV}^{2}$
(in Heaviside-Lorentz units), and
$1\, \mathrm{Mpc} \approx 2 \times 10^{38} \, \mathrm{GeV}^{-1}$.
Moreover, we use Greek letters for the spacetime indices $\mu, \nu = 0,
1, 2, 3$, and Latin letters for spatial indices $i, j = 1,2,3$.

\section{Faraday's Law of Induction Outside the Hubble Horizon}
\label{sec:Faraday}

The magnetic field scaling of~(\ref{B2a6H2}) can be simply understood
from Faraday's law of induction.
In this section we present general arguments that capture the essence
of the physics without specifying the details of the vector field theory.

The electric and magnetic fields measured by a comoving observer with
4-velocity~$u^\mu$ ($u^i=0$, $u_\mu u^\mu = -1$) 
are given by
\begin{equation}
 E_{\mu} = u^\nu F_{\mu \nu} ,
\quad
B_{\mu} = \frac{1}{2} \eta_{\mu \nu \rho \sigma} u^\sigma F^{\nu \rho},
\label{EBformal}
\end{equation}
where $F_{\mu \nu} = \partial_{\mu} A_{\nu} - \partial_\nu A_\mu$,
and $\eta_{\mu \nu \rho \sigma}$ is a totally antisymmetric tensor
with $\eta_{0123} = -\sqrt{-g}$.

Throughout this paper we fix the metric to a flat
FRW,\footnote{Gravitational backreaction on the metric from the gauge
field will be discussed later on.}
\begin{equation}
  ds^2 = a(\tau)^2 \left( -d\tau^2 + d\bd{x}^2 \right).
\label{FRW}
\end{equation}
Then Faraday's law of induction follows from the electromagnetic fields' 
definitions~(\ref{EBformal}) as
\begin{equation}
 (a B_i)' = - \hat{\varepsilon}_{ijl} \, \partial_j (a E_l).
\label{Faraday}
\end{equation}
Here, a prime represents a conformal time~$\tau$ derivative, 
$\hat{\varepsilon}_{ijl}$ is totally antisymmetric with
$\hat{\varepsilon}_{123} = 1$,
and a sum over repeated spatial indices is implied irrespective of
their positions.
Integrating both sides of the equation yields
\begin{equation}
 a B_i = - \hat{\varepsilon}_{ijl}
  \int \frac{da}{a} \frac{\partial_j E_l}{H},
  \label{int-Faraday}
\end{equation}
where we have rewritten the $\tau$-integral in terms of the scale
factor~$a$ and Hubble rate~$H = a' / a^2$.

Now let us go to momentum space,
and focus on modes larger than the Hubble length, i.e., 
on comoving wave numbers that satisfy $k < a H$. 
For such wave modes, the time scales of the electric field oscillations
are longer than the Hubble time, and thus the 
integrand of~(\ref{int-Faraday}) can in many cases
be approximated by some power-law function of~$a$.
Hence Faraday's law implies a relation of
\begin{equation}
 \tilde{B} (\tau, k) \sim \frac{k}{a H}  \tilde{E} (\tau,k) + \frac{C(k)}{a},
\label{B-kaHE}
\end{equation}
where $\tilde{B}$ and $\tilde{E}$ are the Fourier components of $B_{i}$
and $E_i$, respectively,
and we have neglected the spatial indices as well as
$\hat{\varepsilon}_{ijl}$ since we are interested
in order-of-magnitude estimates. 
$C$ is a time-independent integration constant. 
Since the electromagnetic field strengths are written in terms of the vector
components as 
\begin{equation}
 E^2 \equiv E^\mu E_\mu = \frac{E_i E_i }{a^2},
  \quad
 B^2 \equiv B^\mu B_\mu = \frac{B_i B_i }{a^2},
\label{EB2tilde}
\end{equation}
one sees from (\ref{B-kaHE}) that the
part of the magnetic field expressed as the integration constant 
undergoes a radiation-like redshifting~(\ref{B2a4}).
However, there is another part which is related to the electric field as
\begin{equation}
 \Delta B^2 \propto \frac{E^2}{(a H)^2} .
\label{2.7}
\end{equation}
This magnetic component grows relative to the electric field in a
decelerating universe, which can be understood as the 
electric fields sourcing the magnetic fields.
In particular when the electric field is strong enough for this magnetic
component to dominate over the integration constant part, 
and further if the electric field redshifts as $E^2 \propto a^{-4}$,
then the magnetic field would evolve in time
as~(\ref{B2a6H2}).\footnote{We have not discussed the cross term between
the two terms of (\ref{B-kaHE}) since it only becomes marginally
important while $k \tilde{E} / aH$ and $C/a$ are comparable to each other.}

In the above discussions we made some rough approximations upon
obtaining~(\ref{B-kaHE}), however we stress that the argument itself
followed directly from the definitions of the electromagnetic fields.
In particular, we have not specified the gauge field action,
and thus the result applies to the standard Maxwell theory,
as well as to modified electromagnetic theories often invoked in
magnetogenesis scenarios.
In the following sections we give more rigorous arguments 
for a certain class of gauge field theories.

Once the gauge field action is specified,
one obtains the (generalized) Amp\`ere-Maxwell law
(e.g.~(\ref{Ampere})), 
which can be integrated to yield an equation similar to 
(\ref{B-kaHE}) but with $E$ and $B$ flipped,
and with some dependence on the details of the action.
This is useful for studying the relation between the electromagnetic
fields in the presence of strong magnetic fields.
However we should also remark that going between cases of
$E^2 \gg B^2$ and $E^2 \ll B^2$ can be more than just flipping the role
of the electric and magnetic fields;
this reflects the fact that the Amp\`ere-Maxwell law depends on the
gauge field action while Faraday's law is independent.

\section{Electromagnetic Fields and Photon Number}
\label{sec:Bogoliubov}

Hereafter we focus on U(1) gauge field theories described by an
effective action of the form
\begin{equation}
 S = -\frac{1}{4} \int d^4 x \sqrt{-g} \, 
 I(\tau)^2 
  F_{\mu \nu}  F^{\mu \nu} ,
\label{Sem}
\end{equation}
with a time-dependent coefficient~$I(\tau)^2$ of the kinetic term.

The standard Maxwell theory corresponds to the case of $I^2 = 1$,
where the action is invariant under a Weyl transformation,
\begin{equation}
 g_{\mu \nu} \to \Omega^2 g_{\mu \nu}, \quad A_{\mu} \to A_{\mu}.
\end{equation}
Hence with a Weyl-flat background metric such as the flat
FRW~(\ref{FRW}), the gauge field is simply a sum of plane waves.

On the other hand when $I^2$ depends on time, the Weyl invariance is
generically violated and thus the gauge field can be excited even
in a flat FRW universe.
The time-dependent coefficient arises, for instance, from the
Weyl anomaly of quantum
electrodynamics~\cite{Dolgov:1993vg,Benevides:2018mwx}. 
Further time dependence may arise from beyond-the-Standard-Model
physics, such as via couplings of the gauge field to (nearly)
homogeneous degrees of freedom such as the inflaton
field~\cite{Ratra:1991bn}; 
such explicit violation of the Weyl invariance has been invoked in most
primordial magnetogenesis models in the literature.

Below we canonically quantize the theory~(\ref{Sem}),
and write down various quantities in terms of time-dependent Bogoliubov
coefficients.
This will be useful for analyzing the redshifting behaviors of
electromagnetic fields, as well as for studying explicit examples in the
following sections.

\subsection{Canonical Quantization}

We decompose the spatial components of the gauge field into
irrotational and incompressible parts,
\begin{equation}
 A_\mu = (A_0, \partial_i S + V_i)
 \quad  \mathrm{with} \quad
  \partial_i V_i = 0.
\end{equation}
$A_0$ is a Lagrange multiplier in~(\ref{Sem}),
and its constraint equation under proper boundary conditions gives
$A_0 = S'$.
This can be used to eliminate both $A_0$ and $S$
from the action to yield, up to surface terms,
\begin{equation}
 S = \frac{1}{2} \int d \tau d^3 x\,
  I(\tau)^2 \,
  ( V_i' \, V_i' - \partial_i V_j  \, \partial_i V_j ).
\label{SofV}
\end{equation}

We promote~$V_i$ to an operator,
\begin{equation}
 V_i(\tau, \boldsymbol{x}) = 
 \sum_{p = 1,2} \int \frac{d^3 k}{(2 \pi)^3} \, \epsilon^{(p)}_i (\boldsymbol{k})
\left\{
e^{i \boldsymbol{k \cdot x}}  a_{\boldsymbol{k}}^{(p)} 
u^{(p)}_{\boldsymbol{k}} (\tau) + 
e^{-i \boldsymbol{k \cdot x}} a_{\boldsymbol{k}}^{\dagger (p)}
u^{*(p)}_{\boldsymbol{k}} (\tau)  
\right\},
\label{Viop}
\end{equation}
where
$\epsilon_i^{(p)}(\boldsymbol{k})$ ($p = 1,2$) are 
two orthonormal polarization vectors satisfying
\begin{equation}
 \epsilon_i^{(p)} (\boldsymbol{k}) \,  k_i = 0, 
\quad
 \epsilon_i^{(p)} (\boldsymbol{k}) \,  \epsilon_i^{(q)} (\boldsymbol{k})  =
 \delta_{pq}.
\label{2.10}
\end{equation}
It follows from these conditions that
\begin{equation}
 \sum_{p = 1,2} \epsilon_i^{(p)} (\boldsymbol{k}) \, 
\epsilon_j^{(p)} (\boldsymbol{k})
 = \delta_{ij} - \frac{k_i k_j}{k^2},
\label{epepsum}
\end{equation}
where $k \equiv \abs{\bd{k}}$.
Unlike the spacetime indices, we do not assume implicit summation
over the polarization index~$(p)$.

The time-independent annihilation and creation operators,
$a_{\boldsymbol{k}}^{(p)}$ and $a_{\boldsymbol{k}}^{\dagger (p)}$,
satisfy the commutation relations:
\begin{equation}
 [ a_{\boldsymbol{k}}^{(p)},\,  a_{\boldsymbol{h}}^{(q)} ] =
 [ a_{\boldsymbol{k}}^{\dagger (p)},\,  a_{\boldsymbol{h}}^{\dagger (q)}
 ] = 0,
\quad 
 [ a_{\boldsymbol{k}}^{(p)},\,  a_{\boldsymbol{h}}^{\dagger (q)} ] = (2
  \pi)^3 \, 
\delta^{pq} \,
\delta^{(3)}  (\boldsymbol{k} - \boldsymbol{h}) .
 \label{eq:commu}
\end{equation}
Moreover, for $V_i$ and its conjugate
momentum obtained from the action
$S = \int d\tau d^3x \mathcal{L}$ of~(\ref{SofV}) as
\begin{equation}
 \Pi_i = \frac{\partial \mathcal{L}}{\partial V_i'} = I^2 V_i',
\end{equation}
we impose commutation relations as
\begin{equation}\label{eq:commu3}
 \begin{split}
 &\left[ V_i(\tau, \boldsymbol{x}),\,  V_j (\tau, \boldsymbol{y}) \right] = 
 \left[ \Pi_i(\tau, \boldsymbol{x}),\,  \Pi_j (\tau, \boldsymbol{y}) \right]
 = 0,
\\
 \left[ V_i(\tau, \boldsymbol{x}),\,  \Pi_j (\tau, \boldsymbol{y})
  \right] =
  & i \delta^{(3)}   (\boldsymbol{x} - \boldsymbol{y})
\left( \delta_{ij} - \frac{\partial_i \partial_j}{\partial_l \partial_l}
  \right)
  = i \sum_{p = 1,2}\int \frac{d^3 k}{(2\pi)^3} \, 
 e^{i\boldsymbol{k\cdot}  (\boldsymbol{x - y})}
\epsilon_i^{(p)} (\boldsymbol{k}) \, \epsilon_j^{(p)} (\boldsymbol{k}),
 \end{split}
\end{equation}
where the second equality in the second line follows
from~(\ref{epepsum}). 

The mode function $u_{\bd{k}}^{(p)}$ obeys the equation of motion: 
\begin{equation}
 u_{\bd{k}}''^{(p)} + 2 \frac{I'}{I} u_{\bd{k}}'^{(p)} + k^2
  u_{\bd{k}}^{(p)} = 0. 
\label{EoM}
\end{equation}
Choosing the polarization vectors such that
$\epsilon^{(p)}_i (\boldsymbol{k}) = \epsilon^{(p)}_i (-\boldsymbol{k})$,
one can check that the commutation relations (\ref{eq:commu}) and
(\ref{eq:commu3}) are equivalent to each other when the mode function 
is independent of the direction of~$\boldsymbol{k}$, i.e., 
\begin{equation}
 u_{\boldsymbol{k}}^{(p)} = u_k^{(p)},
\label{maru3}
\end{equation}
and also obeys the normalization condition,
\begin{equation}
 I^2 \left(
u_k^{(p)} u'^{*(p)}_k - u_k^{*(p)} u'^{(p)}_k
\right) = i.
\label{v'}
\end{equation}

Defining the vacuum state by
\begin{equation}
 a_{\boldsymbol{k}}^{(p)} |0 \rangle = 0
\label{vacuum}
\end{equation}
for $p = 1,2$ and $^{\forall}  \boldsymbol{k}$,
then the correlation functions of the 
electromagnetic fields~(\ref{EBformal}) can be computed,
\begin{equation}\label{EBcorr}
\begin{split}
 \langle 0| E_\mu (\tau, \boldsymbol{x}) E^\mu (\tau, \boldsymbol{y}) |0
 \rangle &=
 \int \frac{d^3 k}{4 \pi k^3}  e^{i\boldsymbol{k\cdot}  (\boldsymbol{x
- y})} \mathcal{P}_E (\tau, k),
 \\
 \langle 0| B_\mu (\tau, \boldsymbol{x}) B^\mu (\tau, \boldsymbol{y}) |0
 \rangle &=
 \int \frac{d^3 k}{4 \pi k^3}  e^{i\boldsymbol{k\cdot}  (\boldsymbol{x
- y})} \mathcal{P}_B (\tau, k),
\end{split}
\end{equation}
where the power spectra are given in terms of the mode functions as
\begin{equation}
 \mathcal{P}_E(k) = \frac{k^3}{2 \pi^2 a^4} 
\sum_{p=1,2} | u'^{(p)}_k |^2,
\quad
 \mathcal{P}_B(k) = \frac{k^5}{2 \pi^2 a^4} 
\sum_{p=1,2} | u_k^{(p)} |^2.
\label{EBpower}
\end{equation}
We occasionally omit the argument~$\tau$, however it should be noted
that the power spectra generically are time-dependent quantities.

\subsection{Bogoliubov Coefficients}

Since the operators
$a_{\boldsymbol{k}}^{(p)}$ and $a_{\boldsymbol{k}}^{\dagger (p)}$ do not
necessarily diagonalize the Hamiltonian under 
the function~$I(\tau)^2$ with a general time dependence,
let us further introduce a set of time-dependent annihilation and
creation operators
(see \cite{Benevides:2018mwx,Grishchuk:1990bj} for similar analyses
applied to cosmological field excitations),
\begin{equation}
  b_{\bd{k}}^{(p)}(\tau)
   = \alpha_{k}^{(p)} (\tau) \, a_{\bd{k}}^{(p)} +
   \beta_{k}^{*(p)} (\tau) \, a_{-\bd{k}}^{\dagger (p)}, 
\quad
  b_{\bd{k}}^{\dagger (p)}(\tau)
   = \alpha_{k}^{*(p)} (\tau) \, a_{\bd{k}}^{\dagger (p)} +
   \beta_{k}^{(p)} (\tau) \, a_{-\bd{k}}^{(p)},
\end{equation}
where $\alpha_{k}^{(p)} (\tau)$ and $\beta_{k}^{(p)} (\tau)$ are
time-dependent Bogoliubov coefficients expressed in terms of the mode
function as
\begin{equation}
 \alpha_k^{(p)} = I
  \left(\sqrt{\frac{k}{2}} \, u_k^{(p)} + \frac{i}{\sqrt{2 k}} \,
   u_k'^{(p)} \right),
  \quad
 \beta_k^{(p)} = I
  \left(\sqrt{\frac{k}{2}} \, u_k^{(p)} - \frac{i}{\sqrt{2 k}} \,
   u_k'^{(p)} \right).
\label{Bogo}
\end{equation}
One can easily check that 
$b_{\bd{k}}^{(p)}$ and $b_{\bd{k}}^{\dagger (p)}$ satisfy
equal-time commutation relations similar to (\ref{eq:commu}) for
$a_{\bd{k}}^{(p)}$ and $a_{\bd{k}}^{\dagger (p)}$, and also
diagonalize the Hamiltonian,
\begin{equation}
 \tilde{H}  = \int d^3 x \left( \Pi_i V_i' - \mathcal{L} \right)
  = \sum_{p=1,2} \int \frac{d^3 k}{(2 \pi)^3} \, 
  k  \left(
b^{\dagger(p)}_{\bd{k}} b^{(p)}_{\bd{k}} 
+ \frac{1}{2}  [ b^{(p)}_{\bd{k}}, b^{\dagger (p)}_{\bd{k}}]
    \right). 
\label{Hamiltonian}
\end{equation}
It follows from the normalization condition~(\ref{v'}) that the 
Bogoliubov coefficients obey
\begin{equation}
 \abs{\alpha_k^{(p)}}^2 - \abs{\beta_k^{(p)}}^2 = 1,
\label{a-minus-b}
\end{equation}
\begin{equation}
 \abs{\beta_k^{(p)}}^2 = \frac{I^2}{2}
  \left(
k \, \abs{ u_k^{(p)} }^2 +  \frac{\abs{ u_k'^{(p)} }^2}{k}
  \right) - \frac{1}{2}.
\label{beta-amp}
\end{equation}
It is also worth noting that for the standard Maxwell theory
where the mode function is a sum of plane waves (cf.~(\ref{uk-Maxwell})),
the amplitudes $ \abs{\alpha_k^{(p)}}$ and $ \abs{\beta_k^{(p)}} $ are
independent of time. 

Now let us suppose
$a_{\bd{k}}^{(p)}$ and $a_{\bd{k}}^{\dagger (p)}$ 
to have initially diagonalized the Hamiltonian, i.e., $\beta_k^{(p)} =0$
in the distant past,
and that the system was initially in the vacuum state~(\ref{vacuum}).
However, the photons will eventually be produced 
due to the time-dependent background described by~$I(\tau)^2$,
and the number of photons with polarization~$p$ 
per unit six-dimensional phase volume is computed as
\begin{equation}
 \frac{\langle 0| b^{\dagger(p)}_{\bd{k}} b^{(p)}_{\bd{k}} |0
  \rangle}{V}    = \abs{\beta_k^{(p)}}^2,
\end{equation}
where $V$ is the comoving spatial volume,
\begin{equation}
 V \equiv \int d^3 x = (2 \pi)^3 \delta^{(3)} (\bd{0}).
\label{Volume}
\end{equation}
For instance, magnetogenesis models that give rise to
coherent magnetic fields with comoving correlation length of~$k^{-1}$
would create a large number of photons with momentum~$k$, thus yield
$\abs{\beta_k^{(p)}}^2  \gg 1$.  

In terms of the Bogoliubov coefficients,
the electromagnetic spectra (\ref{EBpower}) are written as
\begin{equation}
 \mathcal{P}_E(k) = \frac{k^4}{4 \pi^2 a^4 I^2} 
 \sum_{p=1,2}
 | \alpha_k^{(p)} - \beta_k^{(p)} |^2,
\quad
 \mathcal{P}_B(k) = \frac{k^4}{4 \pi^2 a^4 I^2} 
 \sum_{p=1,2}
 | \alpha_k^{(p)} + \beta_k^{(p)} |^2.
\label{EBpower-ab}
\end{equation}
Here, using (\ref{a-minus-b}), it can be checked that
\begin{equation}
 | \alpha_k^{(p)} \mp \beta_k^{(p)} |^2
  = 1 + 2 \, \abs{\beta_k^{(p)}}^2
  \mp 2 \, \abs{\beta_k^{(p)}} \sqrt{1 + \abs{\beta_k^{(p)}}^2}
  \, \cos \left\{ \arg (\alpha_k^{(p)} \, \beta_k^{(p)*}) \right\},
\label{alphapmbeta}
\end{equation}
which allows the electromagnetic spectra to be expressed in terms of the 
photon number density~$\abs{\beta_k^{(p)}}^2$, and the relative phase
between $\alpha_k^{(p)}$ and $\beta_k^{(p)}$.

The energy density of the gauge field can be obtained as the vacuum
expectation value of the Hamiltonian~(\ref{Hamiltonian}) divided by the
spatial volume, 
\begin{equation}\label{rhoEM}
\begin{split}
 \rho_{A} = \frac{\langle 0| \tilde{H} |0 \rangle}{a^4 V}
  &= \frac{1}{a^4}   \int \frac{d^3k}{(2 \pi)^3} \, k \sum_{p=1,2}
 \left( \abs{\beta_k^{(p)}}^2 + \frac{1}{2} \right)
 \\
 &= \frac{I^2}{2} \int \frac{dk}{k}
  \left\{ \mathcal{P}_E (k) + \mathcal{P}_B (k) \right\},
\end{split}
\end{equation}
where the second line is written in terms of the electromagnetic
power spectra.
In the first line, the $1/2$ inside the parentheses is the zero-point
energy and can be removed by a normal ordering.
(Although, when $\abs{\beta_k^{(p)}}^2 \gg 1$, the zero-point energy is
anyway tiny compared to the total $\rho_A$.)
When $I^2$ is constant and thus 
the photon density $\abs{\beta_k^{(p)}}^2$ is conserved,
one clearly sees that the gauge field density,\footnote{If the
$k$-integral is cut off at some~$k_{\mathrm{UV}}$, in this paragraph we are assuming
$k_{\mathrm{UV}}$ to be time-independent.} 
and the sum of the electric and magnetic power, both redshift as
$\propto a^{-4}$. 
However we stress that this is not necessarily the case for the 
individual electric and magnetic power, as we will explicitly see below.

\subsection{Hierarchical Electromagnetic Power Spectra}

Many Weyl symmetry-breaking models of magnetogenesis
produce much stronger electric fields compared to magnetic fields, or
vice versa.\footnote{This is analogous to the squeezing of inflaton and
graviton fluctuations during inflation~\cite{Grishchuk:1990bj}.}
In terms of the expression~(\ref{alphapmbeta}), 
such a situation with a hierarchy between the electromagnetic fields
is described as the case of
$\abs{\beta_k^{(p)}}^2 \gg 1$
with 
$\arg (\alpha_k^{(p)} \, \beta_k^{(p)*}) \simeq 0, \pm \pi, \pm 2 \pi,
\cdots$.

To see this more clearly, let us write the relative phase as 
\begin{equation}
 \arg (\alpha_k^{(p)} \, \beta_k^{(p)*}) 
 \equiv \pi  + \theta_k^{(p)}.
\label{theta_k}
\end{equation}
One can check that when
\begin{equation}
 \frac{1}{\abs{\beta_k^{(p)}}^2} \ll \abs{\theta_k^{(p)}}  \ll 1
\label{cond}
\end{equation}
is satisfied, then (\ref{alphapmbeta}) is approximated by 
\begin{equation}
    | \alpha_k^{(p)} - \beta_k^{(p)} |^2 \simeq
     4  \, \abs{\beta_k^{(p)}}^2,
\quad
   | \alpha_k^{(p)} + \beta_k^{(p)} |^2 \simeq
   (\theta_k^{(p)})^2 \,  \abs{\beta_k^{(p)}}^2.
\label{apmb}
\end{equation}
This yields 
\begin{equation}
 \mathcal{P}_E(k) \simeq \frac{k^4}{4 \pi^2 a^4 I^2} 
 \sum_{p=1,2} 4  \, \abs{\beta_k^{(p)}}^2,
\quad
 \mathcal{P}_B(k) \simeq \frac{k^4}{4 \pi^2 a^4 I^2} 
 \sum_{p=1,2} (\theta_k^{(p)})^2 \,  \abs{\beta_k^{(p)}}^2,
\label{apmb_spectra}
\end{equation}
describing a much stronger electric field strength compared to the magnetic.
Cases where the magnetic field is stronger can similarly be described
by $\arg (\alpha_k^{(p)} \, \beta_k^{(p)*}) $
being close to~$0$.

\subsection{Maxwell Theory on Super-Horizon Scales}

For the standard Maxwell theory, i.e. $I^2 = 1$,
the mode function is a sum of plane waves,
\begin{equation}
 u_k^{(p)} = \frac{1}{\sqrt{2 k}}
  \left\{  A_k^{(p)} e^{-i k (\tau - \tau_i)}
  + B_k^{(p)} e^{i k (\tau - \tau_i)}\right\}.
\label{plane_waves}
\end{equation}
Here $\tau_i$ is some arbitrary time,
while $A_k^{(p)}$ and  $B_k^{(p)}$ are time-independent complex numbers
satisfying
$ \abs{A_k^{(p)}}^2 - \abs{B_k^{(p)}}^2 = 1$
as required by the normalization condition~(\ref{v'}).
The time-dependent Bogoliubov coefficients are obtained as
\begin{equation}
 \alpha_k^{(p)} = A_k^{(p)} e^{-i k (\tau - \tau_i)},
  \quad
 \beta_k^{(p)} = B_k^{(p)} e^{i k (\tau - \tau_i)} ,
\label{Bogo-Maxwell}
\end{equation}
yielding
\begin{equation}
 \cos \left\{ \arg (\alpha_k^{(p)} \, \beta_k^{(p)*}) \right\}
  = \cos \left\{ \arg (A_k^{(p)} \, B_k^{(p)*})
  - 2 k (\tau - \tau_i) \right\}.  
\end{equation}
Now, supposing that the FRW universe has a constant equation of
state~$w$ ($\neq -1/3$),
the Hubble rate would scale as $H \propto a^{-3 (w+1)/2}$.
Hence the elapsed conformal time is obtained as
\begin{equation}
 \tau - \tau_i = \int^a_{a_i} \frac{da}{a^2 H}
   = \frac{2}{ 3 w + 1}\left( \frac{1}{a H} - \frac{1}{a_i H_i} \right) ,
\label{elapsed_ct}
\end{equation}
where quantities with the subscript~$i$ are evaluated at~$\tau_i$.
Rewriting as
\begin{equation}
 \arg (A_k^{(p)} \, B_k^{(p)*}) = \pi + \Theta_k^{(p)},
\label{eq:Theta_k}
\end{equation}
(note that $\Theta_k^{(p)}$ is independent of time),
then the phase parameter of~(\ref{theta_k}) is
\begin{equation}
 \theta_k^{(p)} = \Theta_k^{(p)}
  - \frac{4}{ 3 w + 1}  \left( \frac{k}{a H} -
   \frac{k}{a_i H_i} \right),
\label{phase_Maxwell}
\end{equation}
up to the addition of integer multiples of $2 \pi$.

\begin{figure}[t]
 \begin{minipage}{.48\linewidth}
  \begin{center}
 \includegraphics[width=\linewidth]{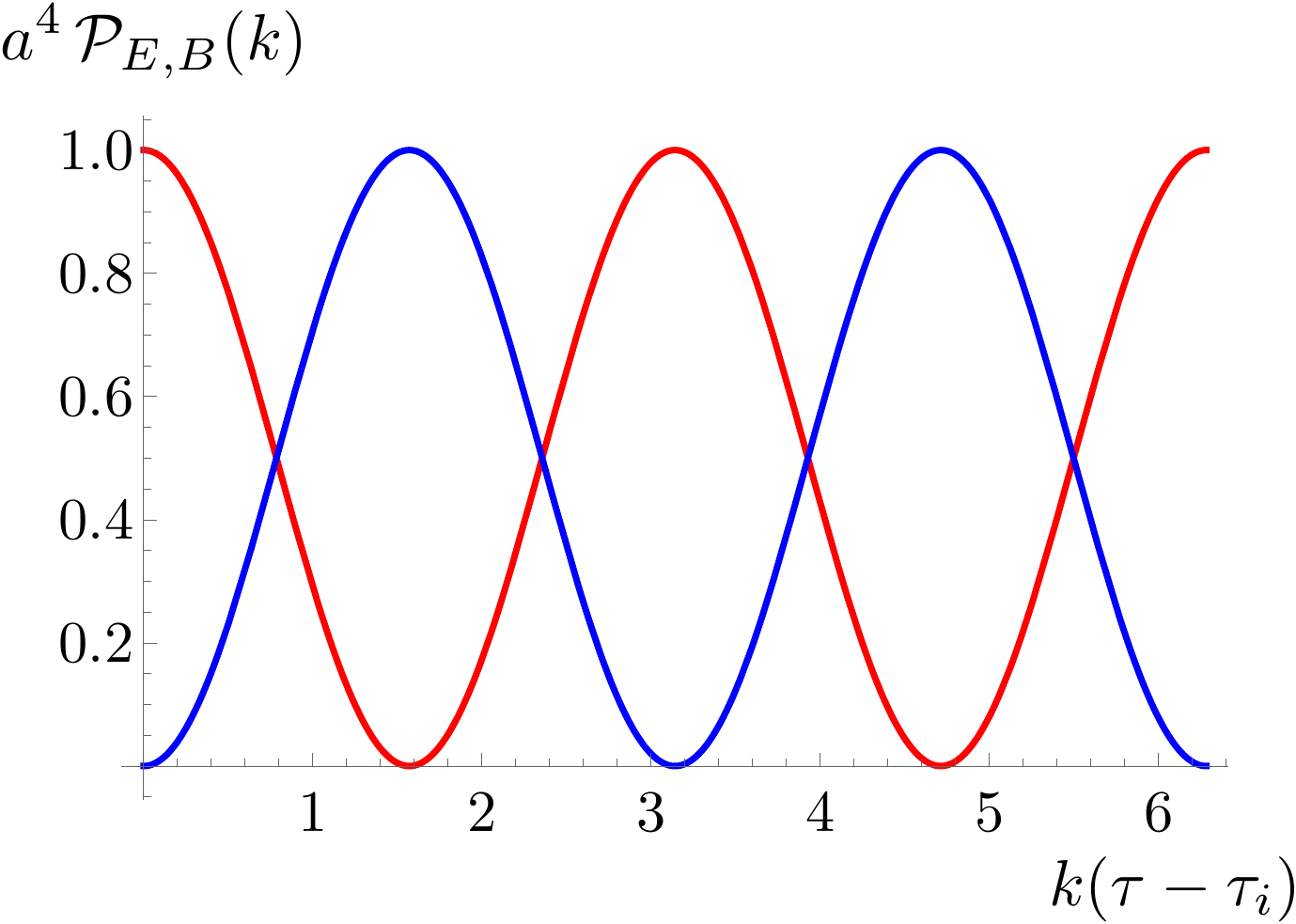}
  \end{center}
 \end{minipage} 
 \begin{minipage}{0.01\linewidth} 
  \begin{center}
  \end{center}
 \end{minipage} 
 \begin{minipage}{.48\linewidth}
  \begin{center}
 \includegraphics[width=\linewidth]{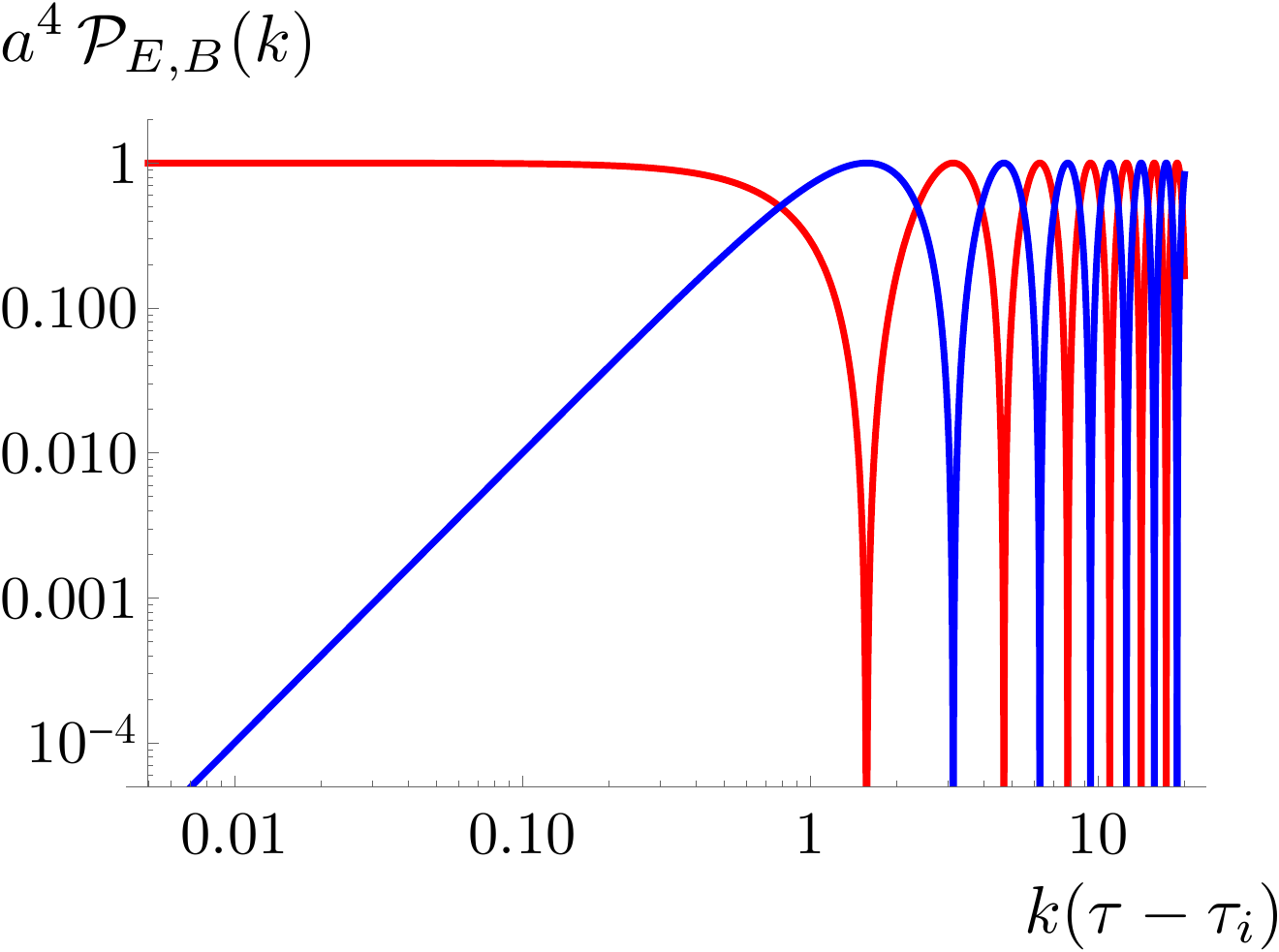}
  \end{center}
 \end{minipage} 
 \caption{Time evolution of electromagnetic fields for the
 standard Maxwell theory, in linear (left panel) and log scales (right
 panel). Shown are the electric (red) and magnetic 
 (blue) power spectra multiplied by $a^4$ and normalized such that their
 oscillation amplitude is unity. The photon density is taken as
 $\abs{\beta_k^{(p)}}^2 \gg 10^2$, and the phase as
 $\Theta_k^{(p)} = 0$. Time is shown in terms of the elapsed conformal
 time in units of $k^{-1}$. When the mode is outside the horizon of a
 decelerating universe, i.e. $k (\tau - \tau_i) \ll 1$, the magnetic
 spectrum grows relative to the electric spectrum which redshifts as
 $\mathcal{P}_E \propto a^{-4}$ (see the text for details).} 
 \label{fig:schematic}
\end{figure}

Let us now consider a situation where there is a hierarchy between the
electric and magnetic power spectra on super-horizon scales.
For this purpose we assume that $\abs{\Theta_k^{(p)}} \ll 1$,
so that $ \theta_k^{(p)}$ is also tiny for
modes satisfying $k \ll a H, a_i H_i$.
Further supposing the photon density $\abs{\beta_k^{(p)}}^2 =
\abs{B_k^{(p)}}^2$ to be large enough to satisfy~(\ref{cond}),
then the super-horizon electromagnetic power spectra are approximately
obtained as 
\begin{equation}\label{3.378}
\begin{split}
 \mathcal{P}_E (k) &\simeq 
  \sum_p \frac{k^4}{\pi^2 a^4} \abs{B_k^{(p)}}^2,
 \\
 \mathcal{P}_B (k) &\simeq
 \sum_p \frac{k^4}{4 \pi^2 a^4} 
 \left\{
\Theta_k^{(p)}
  - \frac{4}{ 3 w + 1}  \left( \frac{k}{a H} -
   \frac{k}{a_i H_i} \right)
  \right\}^2 \abs{B_k^{(p)}}^2 .
\end{split}
\end{equation}
Focusing on the time dependences, 
one sees that the electric power redshifts as~$ \propto a^{-4}$.
The magnetic power, on the other hand, contains a component with a similar
redshifting~$\propto a^{-4}$ (cf.~(\ref{B2a4})) arising
from the $\Theta_k^{(p)}$ and $k/a_i H_i$ terms,
as well as a component with~$\propto a^{-6} H^{-2}$ (cf.~(\ref{B2a6H2}))
arising from the $k/aH$ term.
The former corresponds to the second term in the right hand side 
of~(\ref{B-kaHE}), and the latter corresponds to the first term,
thus manifesting Faraday's law.
If the expansion of the universe is decelerating, i.e. $w > -1/3$,
the magnetic power would eventually be dominated by the 
component with $ \propto a^{-6} H^{-2}$.

In Figure~\ref{fig:schematic} we show the time evolution of
the electromagnetic power spectra for the standard Maxwell theory,
in terms of $k (\tau - \tau_i)$.
Here, the photon density is taken as $\abs{\beta_k^{(p)}}^2 \gg 10^2$, and 
the phase as $\Theta_k^{(p)} = 0$.
The spectra are multiplied by $a^4$, and normalized such that their
oscillation amplitude is unity.
As shown in the left plot, the spectra undergo sinusoidal oscillations
in conformal time.
The super-horizon scaling behaviors of (\ref{3.378}) are easier to see
in the log plot in the right panel.
Here, note that in a decelerating universe ($w>-1/3$), the asymptotic
future corresponds to  $k (\tau - \tau_i) \to \infty$.
One clearly sees from the log plot that when $k (\tau - \tau_i) \ll 1$,
the magnetic field grows relative to the electric field.
On the other hand when $k (\tau - \tau_i) \gg 1$,
the electric and magnetic fields oscillate with similar amplitudes. 

Thus we have explicitly shown for the standard Maxwell
theory that the magnetic power spectrum can scale as (\ref{B2a6H2}) on
super-horizon scales in the presence of stronger electric power.
In the next section we will see how this effect fits within 
magnetic field generation scenarios, by studying a specific model of
inflationary magnetogenesis.

\section{Example: Inflationary Power-Law Magnetogenesis}
\label{sec:power-law}

Let us now study the scaling behaviors of electromagnetic fields
in a specific inflationary magnetogenesis model of the type postulated
in~\cite{Ratra:1991bn} where the $I(\tau)$~function
in the action~(\ref{Sem})
decreases as $\propto a^{-s}$ during inflation, and then becomes
constant after inflation, 
\begin{equation}
  I = 
 \begin{dcases}
     \left(\frac{a_{\mathrm{end}}}{a}\right)^{s} 
       & \text{for $a \leq a_{\mathrm{end}}$,} \\
     1
       & \text{for $a \geq a_{\mathrm{end}}$.}
 \end{dcases}
 \label{power-law-I}
\end{equation}
The subscript ``end'' denotes quantities at the end of inflation.
We take the power~$s$ to be a positive integer, i.e.,
\begin{equation}
 s = 1, 2, 3, \cdots.
\end{equation}
Since $I^2$ does not go below unity in this model,
the gauge kinetic term is never strongly suppressed and thus we do not
worry about strong couplings.

In the following we analyze the cosmological evolution of the
electromagnetic fields during both the inflation and post-inflation epochs using the formalism based on Bogoliubov transformations developed in the previous sections. In appendix \ref{appA}, we reproduce the result by matching directly the classical field across the transition in the long wavelength approximation.
Since the gauge field theory under consideration is
symmetric between the two polarizations,
hereafter we omit the polarization index~$(p)$.

\subsection{Inflationary Magnetogenesis}

During the inflationary epoch $ a \leq a_{\mathrm{end}}$,
we consider the Hubble rate to take a time-independent
value~$H_{\mathrm{inf}}$.
Then the mode function that satisfies the equation of
motion~(\ref{EoM}) and the normalization condition~(\ref{v'}),
as well as approaches a positive frequency solution in the asymptotic
past (i.e. starts from a Bunch-Davies initial condition),
is written in terms of the Hankel function as
\begin{equation}
 u_k =
\frac{1}{2 I} \left( \frac{\pi z}{k } \right)^{\frac{1}{2}}
\, H^{(1)}_{-s + \frac{1}{2}} (z),
    \label{uk_inf}
\end{equation}
up to an unphysical phase. Here the variable~$z$ is defined as
\begin{equation}
 z \equiv \frac{k}{a H_{\mathrm{inf}}}.
\end{equation}
The time-dependent Bogoliubov
coefficients~(\ref{Bogo}) are thus obtained as
\begin{equation}
 \alpha_k =
\left( \frac{\pi z}{8 } \right)^{\frac{1}{2}}
 \left\{
  H^{(1)}_{-s + \frac{1}{2} } (z)
  - i H^{(1)}_{- s -\frac{1}{2} } (z)
	 \right\},
 \quad
  \beta_k =
\left( \frac{\pi z}{8 } \right)^{\frac{1}{2}}
 \left\{
  H^{(1)}_{-s + \frac{1}{2} } (z)
  + i H^{(1)}_{-s -\frac{1}{2} } (z)
\right\}.
\label{Bogo-inf}
\end{equation}

The real and imaginary parts of the Hankel functions are 
respectively the Bessel functions of the first and second kinds,
\begin{equation}
 H_{\nu}^{(1)} (z) = J_{\nu} (z) + i Y_{\nu} (z),
\end{equation}
where $\nu = -s \pm \frac{1}{2}$. 
In the super-horizon limit, i.e. $z \to 0$,
these asymptote to 
(noting that $s$ is a positive integer)~\cite{olver2010nist}, 
\begin{equation}
 J_{\nu} (z) \simeq \frac{1}{ \Gamma (\nu + 1)}
  \left(\frac{z}{2}\right)^\nu, 
\quad
Y_{\nu}  (z) \simeq - \frac{\Gamma (\nu)}{\pi }
\left( \frac{z}{2}  \right)^{-\nu}.
\end{equation}
Using these expressions, one can compute the photon number density as
\begin{equation}
  \abs{\beta_k}^2 \simeq  \frac{\Gamma (s + \frac{1}{2})^2}{4 \pi }
  \left(\frac{2}{z}\right)^{2 s},
\label{beta_inf-super}
\end{equation}
and the phase parameter defined in~(\ref{theta_k}) as,
up to the addition of integer multiples of~$2 \pi$,
\begin{equation}
 \theta_k \simeq - \frac{z}{s - \frac{1}{2}}.
\label{theta_inf-super}
\end{equation}
As the wave mode goes well outside the horizon,
these quantities go as $\abs{\beta_k}^2 \to \infty$ and
$ \theta_k \to 0 $, while satisfying the condition~(\ref{cond}).
Hence one can use the approximation~(\ref{apmb_spectra}) to obtain
the electromagnetic power spectra on super-horizon scales $k \ll a
H_{\mathrm{inf}}$,
\begin{equation}\label{EBpower-inf-super}
\begin{split}
 \mathcal{P}_E (k) &\simeq
 \frac{8 \, \Gamma (s + \frac{1}{2})^2}{\pi^3 } \frac{H_{\mathrm{inf}}^4}{I^2}
 \left(\frac{k}{2 a H_{\mathrm{inf}}}\right)^{-2 (s-2)},
 \\
 \mathcal{P}_B (k) &\simeq 
 \frac{8 \, \Gamma (s - \frac{1}{2})^2}{\pi^3 } \frac{H_{\mathrm{inf}}^4}{I^2}
 \left(\frac{k}{2 a H_{\mathrm{inf}}}\right)^{-2 (s-3)}.
\end{split}
\end{equation}
The two spectra are related via
$ \mathcal{P}_B \simeq (2 s - 1)^{-2}
( k / aH_{\mathrm{inf}} )^2 \mathcal{P}_E$,
which is a manifestation of (\ref{B-kaHE}) implied by Faraday's law.

\subsection{After Inflationary Magnetogenesis}

The universe after inflation stays cold until 
its dominant energy component turns into heat;
we refer to this time when the universe thermalizes as reheating.
During the epoch between the end of inflation and reheating,
let us suppose charged particles to be nonexistent, and also 
the universe to expand with some constant
equation of state~$w$ ($ > -1/3$ such that the expansion decelerates).

Such a post-inflationary expansion can be supported by, for instance,
an inflaton field coherently oscillating about its potential minimum.
If the oscillation is (mostly) along a potential of $V \propto \phi^n$,
the equation of state averaged over the oscillations would be
$w = (n-2)/(n+2)$~\cite{Turner:1983he}.
In this picture, reheating would be induced by the decay of the
inflaton.\footnote{As the oscillation amplitude decreases, eventually,
the potential would likely be dominated by a quadratic term and thus $w$
approaches~$0$. However for simplicity, we
consider $w$ to be constant all the way until reheating.}

\subsubsection*{Between Inflation and Reheating}

We have assumed in~(\ref{power-law-I}) that the standard Maxwell theory
is recovered at the end of inflation.
(Strictly speaking, even within the Standard Model, virtual charged
particles in the loops yield an anomalous dependence of the effective
action for quantum electrodynamics on~$a$, and thus $I$ is not a
constant. However we ignore this since it has little effect on
gauge field excitation~\cite{Benevides:2018mwx}.)
Hence during the cold stage between the end of inflation and reheating,
the gauge field would follow the Maxwell equation in vacuum,
i.e., the equation of motion~(\ref{EoM}) with $I = 1$,
whose solution is given by
\begin{equation}
 u_k = \frac{1}{\sqrt{2 k}}
  \left\{  \alpha_{k} (\tau_{\mathrm{end}})\,
   e^{-i k (\tau - \tau_{\mathrm{end}})} 
   + \beta_{k} (\tau_{\mathrm{end}})\,
   e^{i k (\tau - \tau_{\mathrm{end}})} \right\}.
\label{uk-Maxwell}
\end{equation}
This expression corresponds to~(\ref{plane_waves}) with the choice of
$\tau_i = \tau_{\mathrm{end}}$,
where the coefficients of the positive and negative frequency
solutions are fixed by requiring the Bogoliubov coefficients
during inflation~(\ref{Bogo-inf})
and after~(\ref{Bogo-Maxwell}) to match at~$\tau_{\mathrm{end}}$.
This is equivalent to matching $u_k$ and 
$u_k'$ in the two epochs at the end of inflation.\footnote{The toy model
under consideration involves a sudden 
jump at the end of inflation in the time derivatives of the
$I$~function as well as the Hubble rate~$H$.
Hence depending on whether one chooses to connect $u_k'$ or $(I u_k)'$ or
something else, different results can be obtained.
Here we choose to match the Bogoliubov coefficients since they are
directly related to physical quantities.
We have also verified this procedure by introducing smooth
interpolation for $I$ and $H$ between the two epochs, and numerically
solving the gauge field's equation of motion;
the numerical results agree well with our analytic expressions
(\ref{EBpower-inf-super}) and (\ref{EBpower-post-super})
respectively in the asymptotic regimes $a \ll a_{\mathrm{end}}$
and $a \gg a_{\mathrm{end}}$.}

For wave modes that have exited the horizon during inflation,
the phase parameter in the post-inflation epoch is obtained from
(\ref{phase_Maxwell}) and (\ref{theta_inf-super}) as
\begin{equation}
 \theta_k \simeq - \frac{2}{ 2 s - 1} \, 
\frac{k}{a_{\mathrm{end}} H_{\mathrm{inf}}}
\left\{
1  + \frac{4s -2}{ 3 w + 1}
  \left( \frac{a_{\mathrm{end}} H_{\mathrm{inf}}}{a H} -  1 \right)
\right\},
\end{equation}
whose amplitude monotonically increases in time.
The photon number density $\abs{\beta_k}^2$, which is now
time-independent, is obtained by evaluating
(\ref{beta_inf-super}) at the end of inflation.
The condition~(\ref{cond}) continues to be satisfied
while the mode is well outside the horizon, i.e. $k \ll a H$, 
and thus from (\ref{apmb_spectra}) one can obtain the
electromagnetic power spectra as\footnote{If one allows for a 
general post-inflation expansion history instead of assuming a
constant~$w$, 
then in (\ref{EBpower-post-super}), the final parentheses of
$\mathcal{P}_B$ is replaced by
\begin{equation}
\left\{1  + \frac{4s - 2}{ 3 w + 1}
  \left( \frac{a_{\mathrm{end}} H_{\mathrm{inf}}}{a H} -  1 \right)
	  \right\}^2
\, \to \, 
\left\{
1 + (2 s-1) \int^{a}_{a_{\mathrm{end}}} \frac{da}{a}
  \frac{a_{\mathrm{end}} H_{\mathrm{inf}}}{a H}
\right\}^2.
\end{equation}}
\begin{equation}\label{EBpower-post-super}
\begin{split}
 \mathcal{P}_E (k) &\simeq
 \frac{8 \, \Gamma (s + \frac{1}{2})^2}{\pi^3 } H_{\mathrm{inf}}^4
 \left(\frac{k}{2 a_{\mathrm{end}} H_{\mathrm{inf}}}\right)^{-2 (s-2)}
 \left(\frac{a_{\mathrm{end}}}{a}\right)^4,
 \\
 \mathcal{P}_B (k) &\simeq 
 \frac{8 \, \Gamma (s - \frac{1}{2})^2}{\pi^3 } H_{\mathrm{inf}}^4
 \left(\frac{k}{2 a_{\mathrm{end}} H_{\mathrm{inf}}}\right)^{-2 (s-3)}
\left(\frac{a_{\mathrm{end}}}{a}\right)^4
 \left\{1  + \frac{4s - 2}{ 3 w + 1}
  \left( \frac{a_{\mathrm{end}} H_{\mathrm{inf}}}{a H} -  1 \right)
\right\}^2.
\end{split}
\end{equation}
The decelerated expansion of the universe eventually renders
$a_{\mathrm{end}} H_{\mathrm{inf}} \gg a H $, then
the relation between the electromagnetic power becomes
$ \mathcal{P}_B \simeq (2/3 w+1)^2 ( k / aH )^2 \mathcal{P}_E$,
being compatible with~(\ref{B-kaHE}) which follows from Faraday's law.
Here, it is also important to note that while the electric power
redshifts as $\mathcal{P}_E \propto a^{-4}$,
the magnetic power scales\footnote{If the standard Maxwell theory is
recovered during 
inflation instead of at the very end, the magnetic power would initially
redshift as $\mathcal{P}_B \propto a^{-4}$, then some time after
inflation switch to $\propto a^{-6} H^{-2}$.}
as $\mathcal{P}_B \propto a^{-6} H^{-2} \propto a^{3 (w-1)}$.

\subsubsection*{After Reheating}

Upon reheating, 
the conductivity of the universe becomes high,
and thus the electric fields are shorted out while the magnetic
flux is frozen in.
Hence we consider large-scale magnetic fields after reheating to
redshift as $\mathcal{P}_B \propto a^{-4}$ until today.

The magnetic power spectrum in the present universe is thus obtained as,
for wave modes that are outside the horizon at the time of
reheating,\footnote{Reheating happens before Big Bang
Nucleosynthesis (BBN), and since the comoving Hubble radius at
the beginning of BBN is of
$a_0/(a_{\mathrm{BBN}} H_{\mathrm{BBN}}) \sim 10\, \mathrm{pc}$,
the result~(\ref{Bpower-today-s}) applies at least for
wave numbers satisfying $k/ a_0 < (10 \, \mathrm{pc})^{-1}$.}
\begin{equation}\label{Bpower-today-s}
\begin{split}
 \mathcal{P}_{B0} (k)
 &= \mathcal{P}_{B\, \mathrm{reh}} (k)
 \left(\frac{a_{\mathrm{reh}}}{a_0}\right)^4
 \\
 &\simeq 
 \frac{8 \, \Gamma (s - \frac{1}{2})^2}{\pi^3 } H_{\mathrm{inf}}^4
 \left(\frac{k}{2 a_{\mathrm{end}} H_{\mathrm{inf}}}\right)^{-2 (s-3)}
 \left(\frac{a_{\mathrm{end}}}{a_0}\right)^4
 \left\{1  + \frac{4s - 2}{ 3 w + 1}
  \left( \frac{a_{\mathrm{end}} H_{\mathrm{inf}}}{a_{\mathrm{reh}}
   H_{\mathrm{reh}}} -  1 \right) 
\right\}^2,
\end{split}
\end{equation}
where the subscript ``reh'' is used to describe quantities upon
reheating, and ``$0$'' for today.
The enhancement factor of
\begin{equation}
 \frac{a_{\mathrm{end}} H_{\mathrm{inf}}}{a_{\mathrm{reh}}
  H_{\mathrm{reh}}}
  = \left( \frac{H_{\mathrm{inf}}}{H_{\mathrm{reh}}}  \right)^{\frac{3 w+1}{3 w+3}}
\label{inf-reh-fac}
\end{equation}
inside the parentheses represents the effect of the electromagnetic
induction during the epoch between inflation and reheating.
This would be missed if one were to assume the magnetic power to 
redshift as $a^{-4}$ right from the end of inflation, as has been done
in most previous works.
The enhancement factor becomes particularly large when there is a
hierarchy between the inflation and reheating scales.
The scale of inflation is bounded from above by the current observational
limit on primordial gravitational waves as 
$H_{\mathrm{inf}} \lesssim 10^{14}\, \mathrm{GeV}$~\cite{Akrami:2018odb},
while the reheating temperature needs to be higher than about $5 \,
\mathrm{MeV}$ in order not to spoil BBN~\cite{Hannestad:2004px}, setting
a lower bound on the reheating 
scale as $H_{\mathrm{reh}} \gtrsim 10^{-23}\, \mathrm{GeV}$.
Hence the ratio between the inflation and reheating scales can in
principle be as large as 
$H_{\mathrm{inf}} / H_{\mathrm{reh}} \lesssim 10^{37} $,
and the post-inflationary induction would significantly
impact the final magnetic field amplitude.
The effect is maximized for a stiff equation of state $ w \gg 1$,
for which the factor of~(\ref{inf-reh-fac}) can be as large as~$10^{37}$.
If $w = 1/3$, which is the case we will mainly consider in the example
in the next subsection, the factor can be up to~$10^{18}$. 
Even with a pressureless state $w = 0$, the factor can be as large as~$10^{12}$.

\subsection{Intergalactic Magnetic Fields from High-Scale Inflation}

To demonstrate the importance of the post-inflationary
induction, let us present an example
where the femto-Gauss intergalactic magnetic fields as suggested by 
recent gamma ray observations
are produced from inflationary magnetogenesis with a high
inflation scale and low reheat temperature.

The example is given by the model of (\ref{power-law-I}) with a power
\begin{equation}
 s = 2,
\end{equation}
which produces a $k$-independent electric power spectrum,
cf.~(\ref{EBpower-inf-super}).
The gauge field's energy density~(\ref{rhoEM}) during inflation is
dominated by the scale-invariant electric power, which is roughly of order
\begin{equation}
 \rho_{A} \sim H_{\mathrm{inf}}^4
  \,  \log \left(\frac{a H_{\mathrm{inf}}}{k_{\mathrm{IR}}}\right).
 \label{rhoEM_inf-s2}
\end{equation}
Here, upon carrying out the $k$-integral in~(\ref{rhoEM}),
we have introduced a UV cutoff and set it to the mode exiting the
horizon, i.e. $k_{\mathrm{UV}} \sim a H_{\mathrm{inf}}$,
since for higher $k$~modes the gauge field fluctuations have not yet
become classical and thus their contributions to the energy density
should be renormalized.\footnote{By `becoming classical,' we mean that
the classical volume of the space spanned by the gauge field fluctuation
and its conjugate momentum becomes much larger than their quantum
uncertainty.
See~\cite{Green:2015fss,Benevides:2018mwx,Maldacena:2015bha}
for detailed analyses.} 
We have also introduced an IR cutoff~$k_\mathrm{IR}$; considering it to 
be the wave mode that exited the horizon at the beginning of inflation,
the factor~$\log (a H_{\mathrm{inf}} / k_{\mathrm{IR}})$
corresponds to the number of elapsed inflationary $e$-folds~$\mathcal{N}$.
Here, from the observational limit 
$H_{\mathrm{inf}} \lesssim 10^{14}\, \mathrm{GeV} $,
the ratio between the gauge field density~(\ref{rhoEM_inf-s2}) 
and the total density of the universe
$\rho_{\mathrm{tot}} = 3 M_p^2 H_{\mathrm{inf}}^2$
is bounded as
$\rho_{A} / \rho_{\mathrm{tot}} \lesssim 10^{-9} \mathcal{N}$,
being much smaller than the amplitude of the
curvature perturbation $\zeta \sim 10^{-5}$ measured on CMB scales
(unless the inflationary period is extraordinarily long).
Thus the effect of the excited gauge field on the cosmological
perturbations\footnote{The gauge field sources curvature perturbations
roughly of $\zeta_A \sim \rho_A / (\epsilon
\rho_{\mathrm{tot}})$, where $\epsilon = -H' / (a H^2)$ is the rate of
change of the Hubble parameter.
See e.g.~\cite{Ferreira:2014hma} for detailed analyses of CMB
constraints.}
and the inflationary background is negligible.

However we should also remark that, depending on the post-inflationary
equation of state~$w$, the gauge field's backreaction may become
non-negligible after inflation.
Here, recall that once the standard Maxwell theory is recovered,
the gauge field density redshifts as radiation,
i.e. $\rho_{A} \propto a^{-4}$. 
Hence if $w \leq 1/3$, its ratio to the total density~$\rho_A /
\rho_{\mathrm{tot}}$ does not increase in time.
However if $w > 1/3$, the ratio would grow and thus one needs to verify
whether the backreaction becomes significant.

\begin{figure}[t]
  \begin{center}
  \begin{center}
  \includegraphics[width=0.58\linewidth]{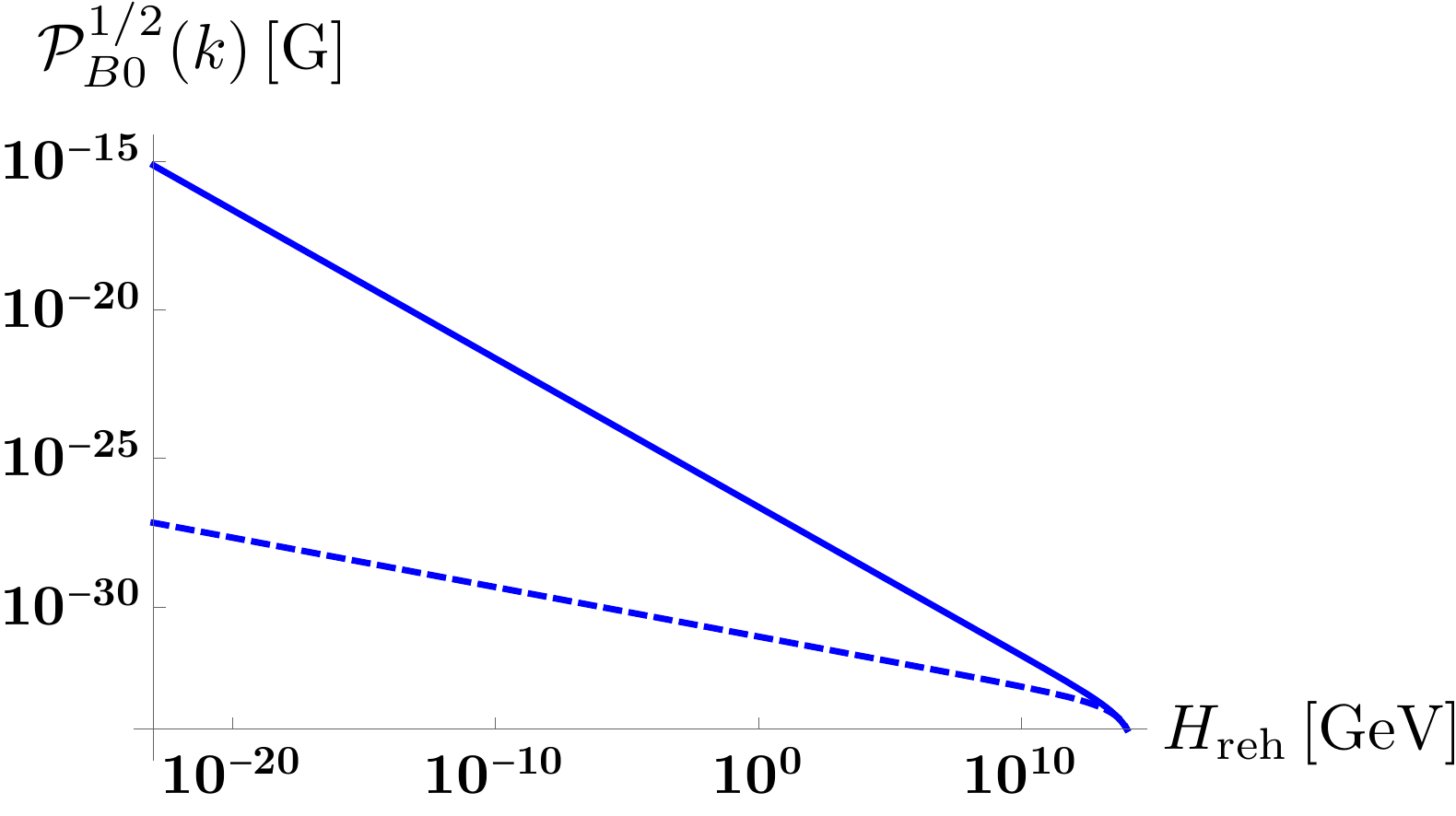}
  \end{center}
  \caption{Magnetic field strength today on $k/a_0 = (1 \,
   \mathrm{Mpc})^{-1}$, generated by the inflationary magnetogenesis
   model~(\ref{power-law-I}) with $s = 2$. The inflation scale is fixed
   to $H_{\mathrm{inf}} = 10^{14}\, \mathrm{GeV}$, and the field
   strength is 
   shown as a function of the  Hubble scale at reheating. The
   post-inflationary equation of state is taken as $w = 0$ (dashed line)
   and $w = 1/3$ (solid).}
  \label{fig:PB0}
  \end{center}
\end{figure}

The magnetic field strength today is obtained by
substituting $s = 2$ into (\ref{Bpower-today-s}),
and let us suppose that $ a_{\mathrm{end}} H_{\mathrm{inf}} \gg
a_{\mathrm{reh}} H_{\mathrm{reh}} $, namely,
that the universe thermalizes well after inflation ends.
Considering the entropy of the universe to be conserved since
reheating, the redshift and energy scale of reheating are related by
(supposing the Standard Model degrees of freedom),
\begin{equation}
 \frac{a_0}{a_{\mathrm{reh}}}
  \approx 3 \times 10^{10}
  \left( \frac{H_{\mathrm{reh}}}{10^{-23}\, \mathrm{GeV}}  \right)^{1/2}.
\label{z-reh}
\end{equation}
Also using (\ref{inf-reh-fac}),
the magnetic field spectrum on large scales is obtained as
\begin{equation}
 \mathcal{P}_{B0} (k) \sim
  \frac{(10^{-33}\, \mathrm{G})^2}{(3 w + 1)^2}
  \left(\frac{k}{a_0} \, \mathrm{Mpc} \right)^2
  \left( \frac{H_{\mathrm{inf}}}{10^{14}\, \mathrm{GeV}} \right)
  \left(\frac{H_{\mathrm{inf}}}{H_{\mathrm{reh}}}\right)^{\frac{9
  w+1}{3w+3}}. 
\label{eq:4.20}
\end{equation}
From this expression one sees that the magnetic field strength is
larger for smaller length scales, higher inflation scales, 
and if $ w > -1/9$,
for larger $H_{\mathrm{inf}} / H_{\mathrm{reh}}$~ratios.
In the case with the largest possible hierarchy
between the inflation and reheating scales\footnote{The case $\rho_{\mathrm{inf}}^{1/4} = 10^{16}$GeV ($H_{\mathrm{inf}}\sim 10^{14}$GeV) and $s=2$ is within the region where backreaction and anisotropy constraints are satisfied \cite{Ferreira:2014hma}.}, i.e., 
$H_{\mathrm{inf}} = 10^{14}\, \mathrm{GeV}$ and 
$H_{\mathrm{reh}} = 10^{-23}\, \mathrm{GeV}$,
the magnetic field strength
on the wave number $k/a_0 = (1 \, \mathrm{Mpc})^{-1}$ is
$\mathcal{P}_{B0}^{1/2} \sim 10^{-27}\, \mathrm{G}$ for $w = 0$, and
$\mathcal{P}_{B0}^{1/2} \sim 10^{-15}\, \mathrm{G}$ for $w = 1/3$.
For the same parameters but with a higher reheating scale 
$H_{\mathrm{reh}} = 10^{-12}\, \mathrm{GeV}$ (corresponding to
a temperature of $T_{\mathrm{reh}} \sim 1 \, \mathrm{TeV}$),
then $\mathcal{P}_{B0}^{1/2} \sim 10^{-21}\, \mathrm{G}$ for $w = 1/3$.
In Figure~\ref{fig:PB0} we plot the magnetic field strength as a
function of~$H_{\mathrm{reh}}$, for $k/a_0 = (1 \, \mathrm{Mpc})^{-1}$
and $H_{\mathrm{inf}} = 10^{14}\, \mathrm{GeV}$. The dashed line shows
the case of $w = 0$, while the solid line is for $w = 1/3$. 
The lines are seen to bend at $H_{\mathrm{reh}} \gtrsim 10^{13}\,
\mathrm{GeV}$; here reheating happens soon after inflation and hence
there is not enough time for the induction effect to become important, 
namely, the second term inside the $\{\, \}$ parentheses
of~(\ref{Bpower-today-s}) is not much greater unity
and thus the result deviates from the approximation~(\ref{eq:4.20}).
The field strength basically increases with~$w$, however for $ w > 1/3$, 
the post-inflation backreaction may become non-negligible
as discussed above.\footnote{An equation of state of $ w > 1/3$
can also blue-tilt the primordial gravitational wave
spectrum~\cite{Figueroa:2019paj}. It would be interesting to study the
possibility of probing~$w$ from a joint analysis of the magnetic
fields and gravitational waves.}

Thus by properly taking into account electromagnetic induction after
inflation, we have shown that the simple inflationary magnetogenesis
model~(\ref{power-law-I}) with $s = 2$ is capable of creating femto-Gauss
intergalactic magnetic fields on Mpc scales, given a high-scale
inflation $H_{\mathrm{inf}} = 10^{14}\, \mathrm{GeV}$ and
low reheating $H_{\mathrm{reh}} = 10^{-23}\, \mathrm{GeV}$,
with the two periods connected by an equation of state~$ w = 1/3$. 
Here we stress that the equation of state $w=1/3$ of this scenario is
not due to charged relativistic particles, but instead should be
realized by some substance without charge such as an oscillating
inflaton condensate.

\section{Model-Independent Constraints}
\label{sec:MIC}

We have shown in the previous sections that if primordial magnetogenesis
creates stronger electric fields than magnetic fields,
then even after the standard Maxwell theory is recovered, 
the electromagnetic spectra on super-horizon scales can be related by
\begin{equation}
 \mathcal{P}_{B} (k) \sim
  \left( \frac{k}{a H}  \right)^2
  \mathcal{P}_{E} (k),
\label{gen-assump1}
\end{equation}
yielding the magnetic scaling $\mathcal{P}_B \propto a^{-6 } H^{-2}$
instead of a radiation-like redshifting.
In this section we derive generic bounds on primordial magnetic fields
with such a behavior, by analyzing the gauge field's
gravitational backreaction.

\subsection{Generic Reheating Bound}
\label{subsec:GRB}

We start by constraining cases where the standard Maxwell theory
is recovered by the time of reheating.
(Thus the Weyl invariance of the gauge field action can 
explicitly be violated even after inflation, as in post-inflationary 
magnetogenesis scenarios~\cite{Kobayashi:2014sga}, see also~\cite{Fujita:2016qab,Sharma:2017eps}.)
We suppose that coherent electromagnetic fields have been created on
some wave modes that are outside the horizon upon reheating, 
and that right before reheating when the electric fields have not
yet vanished, the power spectra satisfy 
the relation~(\ref{gen-assump1}) on the wave modes of interest.

One can read off from the second line of~(\ref{rhoEM})
that in Maxwell theory ($I^2 = 1$), 
the electric power spectrum with wave number~$k$
contributes to the gauge field's energy density as
$\Delta \rho_A \sim \mathcal{P}_E (k) / 2$, 
given that the spectrum~$\mathcal{P}_E (k)$ is
smooth over a range of $\Delta k \sim k$ so that the integral $\int
dk/k$ can be approximated by an order-unity factor.\footnote{Sharp
features localized to ranges of $\Delta k \ll k$ 
can be produced if rapidly time varying backgrounds give
rise to resonant production of photons. This could in principle provide
a way to evade the constraints in this section.}
Considering that the other contributions to the gauge field density are
non-negative, an inequality of
\begin{equation}
 \rho_{A} \gtrsim \frac{1}{2} \mathcal{P}_{E} (k)
\label{gen-assump2}
\end{equation}
is thus obtained.
We further assume the magnetic power to redshift after reheating as
\begin{equation}
 \mathcal{P}_B(k) \propto a^{-4}
 \quad (a \geq a_{\mathrm{reh}}).
\label{gen-assump3}
\end{equation}

Based on these assumptions, an upper bound on the magnetic spectrum
in the current universe is obtained as
\begin{equation}\label{reh-limit}
\begin{split}
 \mathcal{P}_{B0} (k) &=
 \mathcal{P}_{B\, \mathrm{reh}} (k) \left( \frac{a_{\mathrm{reh}}}{a_0}
 \right)^4
 \\
 &\lesssim
 6 M_p^2 \left( \frac{k}{a_0}\right)^2
  \left( \left. \frac{\rho_{A}}{\rho_{\mathrm{tot}}}
 \right|_{\mathrm{reh}}\right)
  \left( \frac{a_{\mathrm{reh}}}{a_0} \right)^2
 \\
 &\sim
 (10^{-13}\, \mathrm{G})^2 \left( \frac{k}{a_0}\, \mathrm{Mpc} \right)^2
  \left( \left. 10^5  \frac{\rho_{A}}{\rho_{\mathrm{tot}}}
 \right|_{\mathrm{reh}}\right)
 \left( \frac{10^{-23}\, \mathrm{GeV}}{H_{\mathrm{reh}}} \right).
\end{split}
\end{equation}
Here we have used
(\ref{gen-assump3}) in the first line,
then $\rho_{\mathrm{tot}} = 3 M_p^2 H^2$,
(\ref{gen-assump1}), and (\ref{gen-assump2}) to get to 
the second line, and (\ref{z-reh}) for the third line.
The reference value of
$(\rho_{A} / \rho_{\mathrm{tot}})_{\mathrm{reh}} = 10^{-5}$
has been chosen from the amplitude of the large-scale curvature
perturbation $\zeta \sim 10^{-5}$; 
a larger density ratio, in particular if its main contribution is
on CMB scales, would source too large curvature perturbations and
contradict with observations. 
Note also that $H_{\mathrm{reh}} = 10^{-23}\, \mathrm{GeV}$ is the
lowest possible reheating scale compatible with BBN.
Hence the bound~(\ref{reh-limit}) shows that if the electromagnetic
spectra satisfy the relation~(\ref{gen-assump1}) right before reheating,
then the magnetic field strength cannot exceed $10^{-13}\,
\mathrm{G}$ on $\mathrm{Mpc}$ or larger scales today,
otherwise the gauge field fluctuations would spoil the
cosmological perturbations.
In particular, in order to have femto-Gauss magnetic fields on Mpc
scales, the reheating scale should satisfy
$H_{\mathrm{reh}} \lesssim 10^{-18}\, \mathrm{GeV}$, which in terms of
the reheating temperature translates into
$T_{\mathrm{reh}} \lesssim 1\, \mathrm{GeV}$.

Upon deriving the bound~(\ref{reh-limit}), we have only employed
assumptions about times from reheating onward.
In particular, no assumption was made regarding 
cosmology and the gauge field theory in epochs prior to reheating.

\subsection{Less Generic Inflation Bound}

Let us now make some assumptions about the period 
between inflation and reheating, in order to obtain a magnetic field
bound in terms of the inflation scale. 
Hereafter we assume that by the end of inflation, the standard Maxwell
theory is recovered and yields (\ref{gen-assump2})
(hence the following discussions are limited to inflationary
magnetogenesis scenarios).
We further assume that the post-inflationary universe expands with a
constant equation of state~$w$ until reheating,
with the electric field redshifting as $\mathcal{P}_E \propto a^{-4}$
during this period.
As in Section~\ref{subsec:GRB}, we suppose the relation
(\ref{gen-assump1}) to hold right before reheating, and the
magnetic field to redshift as (\ref{gen-assump3}) after reheating. 

Then in a similar way as we derived~(\ref{reh-limit}), but now
considering the backreaction at the end of inflation
(note $H(a_{\mathrm{end}}) = H_{\mathrm{inf}}$),
one can obtain
\begin{equation}
  \mathcal{P}_{B0} (k) \lesssim
 6 M_p^2 \left( \frac{k}{a_0}\right)^2
  \left( \left. \frac{\rho_{A}}{\rho_{\mathrm{tot}}}
	 \right|_{\mathrm{end}}\right) 
 \left( \frac{a_{\mathrm{reh}}}{a_0} \right)^2
  \left(\frac{H_{\mathrm{reh}}}{H_{\mathrm{inf}}}
  \right)^{\frac{2 (-3 w +1)}{3 (w+1)}}.
\label{end-limit}
\end{equation}
The main difference from the reheating bound~(\ref{reh-limit}) is the 
presence of the ratio~$H_{\mathrm{reh}}/H_{\mathrm{inf}}$.
It appears in the bound with a positive (negative) power for
$w$~$<$~($>$)~$1/3$, 
reflecting the fact that the gauge density ratio~$\rho_A /
\rho_{\mathrm{tot}}$ decreases (increases) in time after inflation.
Thus the bound would be equivalent to~(\ref{reh-limit})
if $w = 1/3$ (i.e. radiation-like background), or
$H_{\mathrm{reh}} = H_{\mathrm{inf}}$ (i.e. instantaneous reheating at
the end of inflation).

If, for instance, $w = 0$, then (\ref{end-limit}) can be
rewritten as
\begin{equation}
  \mathcal{P}_{B0} (k) \lesssim
 (10^{-15}\, \mathrm{G})^2 \left( \frac{k}{a_0}\, \mathrm{Mpc} \right)^2
  \left( \left. 10^5 \frac{\rho_{A}}{\rho_{\mathrm{tot}}}
 \right|_{\mathrm{end}}\right)
  \left( \frac{10^{-23}\, \mathrm{GeV}}{H_{\mathrm{reh}}} \right)^{1/3}
  \left( \frac{10^{-16}\, \mathrm{GeV}}{H_{\mathrm{inf}}} \right)^{2/3}.
\label{5.6}
\end{equation}
Hence one finds that for $w = 0$, femto-Gauss magnetic fields can exist
on Mpc scales only if the inflation scale satisfies\footnote{In
\cite{Green:2015fss}, constraints on inflationary  
magnetogenesis were derived for general gauge field theories with a
two-derivative kinetic term,
under the assumption of the post-inflationary redshifting $\mathcal{P}_B
\propto a^{-4} $.
Their bound~(3.20), for instance, is modified 
by instead adopting $\mathcal{P}_B \propto a^{-6} H^{-2}$;
further multiplying by~$10^{-5}$
considering the curvature perturbation, the modified bound matches with
our~(\ref{5.6}).} 
$H_{\mathrm{inf}} \lesssim 10^{-16}\, \mathrm{GeV}$.

\section{Comments on Schwinger Effect}
\label{sec:Schwinger}

The non-radiation-like scaling of the electromagnetic fields 
arises in the presence of a hierarchy between the electric and
magnetic field strengths.
In the previous sections we considered electric
fields much stronger than magnetic fields being produced in the
primordial universe, which then affect the subsequent magnetic field
evolution. Up until the time of reheating, we supposed a cold
universe where charged particles are absent, and hence assumed the
electric fields to survive.
However, if the electric field is strong enough, it can give rise to
Schwinger production of charged
particles~\cite{Sauter:1931zz,Heisenberg:1935qt,Schwinger:1951nm},
which in turn would backreact significantly on the electric fields 
before reheating~\cite{Kobayashi:2014zza}
(see also e.g. \cite{Hayashinaka:2016qqn,Sobol:2018djj,Banyeres:2018aax}).

Studying the fate of strong cosmological electric fields would
require an analysis of the Schwinger process in a curved
spacetime, whose behavior can differ from that in flat space due to the
extra effect from the gravitational background,
as was shown explicitly for de Sitter spacetimes
in e.g.~\cite{Kobayashi:2014zza,Garriga:1994bm,Frob:2014zka}. 
A complete analysis in a generic FRW spacetime is beyond the scope of
this paper; instead we provide here a crude estimate of the impact of
the Schwinger process in a cosmological background, and postulate the
condition under which primordial electric fields are unaffected by the
Schwinger effect.

We again consider the gauge field theory of~(\ref{Sem}), but now
coupled to matter such that the equation of motion of the gauge field
includes a conserved current~$J^\mu$,
\begin{equation}
 \nabla_\mu (I^2 F^{\mu \nu}) = - J^\nu.
\end{equation}
Its spatial component
yields the modified Amp\`ere-Maxwell law,
which in terms of the 
electromagnetic fields reads
(our sign convention follows from the definition~(\ref{EBformal})
with $u^0 > 0$)
\begin{equation}
 \hat{\varepsilon}_{ijl} \, \partial_j B_l =
  \frac{(a I^2 E_i)'}{a I^2} + \frac{a J_i}{I^2}
  = E_i' +
  \left\{ \frac{(a I^2)'}{a I^2} + \frac{a \sigma }{I^2} \right\}E_i.
\label{Ampere}
\end{equation}
Here in the far right hand side,
the current is considered to be carried by particles produced via
the Schwinger process, and thus we have rewritten it using the
conductivity~$\sigma$ introduced as 
\begin{equation}
 J_i = \sigma E_i.
\end{equation}
Let us now assume the first term inside the $\{\, \}$ parentheses to be of
\begin{equation}
 \left| \frac{(a I^2)'}{a I^2} \right| \sim a H,
\end{equation}
as is the case for the standard Maxwell theory ($I^2 = 1$), as well as the
power-law magnetogenesis model in Section~\ref{sec:power-law}.
Then the condition under which the induced current has a negligible
effect on the evolution of the electric field can be read off as
\begin{equation}
 \abs{\sigma} \lesssim I^2 H.
\label{6.5}
\end{equation}

In Minkowski space, the conductivity induced by a background electric
field through the Schwinger pair production is of 
(see e.g. \cite{Anderson:2013ila})
\begin{equation}
 \sigma \sim (t-t_\mathrm{on}) e^3 E  \exp\left( -\frac{\pi m^2}{eE}\right),
\label{sigma-Min}
\end{equation}
where $t$ is time, $t_{\mathrm{on}}$ is when the electric field was
turned on, $E$ is the electric field strength, 
and $m$ and $e$ are respectively the mass and amplitude of the
charge of the produced pairs.
The linear dependence on time reflects the
fact that the produced particles accumulate until their backreaction to
the electric field becomes non-negligible.

On the other hand in a cosmological background,
the expansion of the universe dilutes away the particles produced by
the electric field, thus introducing the time scale~$H^{-1}$.
Hence, supposing that the rate of change of the electric
field is comparable to or smaller than~$H$,
we crudely estimate the induced conductivity in a FRW
background by replacing the elapsed time in the flat space
result~(\ref{sigma-Min}) by the Hubble time,\footnote{We assume the
$I^2$~function to multiply only the photon kinetic term but not the 
photon-matter coupling terms, therefore the induced conductivity
would not explicitly depend on~$I^2$.}
\begin{equation}
 \sigma \sim \frac{e^3 E}{H} \exp\left( -\frac{\pi m^2}{eE}\right).
\label{sigma-FRW}
\end{equation}
Here the electric field strength is understood as
$E = (E_\mu E^\mu)^{1/2}$.
In an inflationary de Sitter space,
the conductivity induced by a time-independent electric field actually 
does take this form in the strong electric field regime $e E \gg H^2$;
while with weak electric fields, gravitational particle production
renders $\sigma$ to take a different form~\cite{Kobayashi:2014zza}.
Since now we are interested in strong primordial electric fields,
let us adopt (\ref{sigma-FRW}) for the moment as the induced
conductivity in a generic FRW universe.

Then the condition~(\ref{6.5}) reads
\begin{equation}
 \frac{e^3 E}{H^2} \exp\left( -\frac{\pi m^2}{eE}\right)  \lesssim I^2.
\label{f-cond}
\end{equation}
This can be understood as an upper bound on the electric field strength
for which the backreaction from the produced particles can be neglected.
In particular if the charged particle is light enough such that
$m^2 \ll e E$, then the bound is simplified to 
\begin{equation}
\frac{ e^3 E}{H^2} \lesssim I^2 ,
\label{6.9}
\end{equation}
which implies that if light charged particles exist in the theory, then
electric fields exceeding the Hubble scale multiplied by~$I^2$
would receive significant backreaction from the Schwinger
process.\footnote{Given that $e^2/I^2 \lesssim 1$, then the
regime affected by the Schwinger process, i.e. $e^3 E \gg I^2 H^2$,
would fall into the strong field regime $e E \gg H^2$ where the
approximation 
(\ref{sigma-FRW}) is expected to hold. This justifies our use of
(\ref{sigma-FRW}) for constraining electric fields.}

When applying the above discussion to the inflationary magnetogenesis
scenario of Section~\ref{sec:power-law} by the substitution
$E \to \mathcal{P}_E(k)^{1/2}$, one can check that
the condition~(\ref{6.9}) for $e \sim 1$ is either saturated or violated
(depending on the value of~$s$)
at some $k$-mode towards the end of inflation.
Moreover in the post-inflation epoch, the condition is strongly violated 
since the ratio $P_E(k)^{1/2} / H^2 \propto a^{1+3 w}$ grows in time. 
Hence our crude estimate suggests that the evolution of the electric
field is affected by the produced light charged particles before
reheating, and thus the magnetic scaling would deviate from $P_B(k)
\propto a^{-6} H^{-2}$. 
The Schwinger production, however, could be avoided if there is some
mechanism in the early universe giving sufficiently large masses to
charged particles.\footnote{In \cite{Kobayashi:2014zza}, Schwinger
effect constraints on inflationary magnetogenesis were derived by
analyzing the Schwinger process during inflation, and assuming
$\mathcal{P}_B \propto a^{-4}$ after inflation.
The constraints can be relaxed in the presence of the post-inflation
induction; however the estimate in this section indicates that even if
the Schwinger process during inflation is negligible, it may become
important afterwards.}

We stress that the analyses in this section
rely on the very rough estimate of the conductivity (\ref{sigma-FRW})
induced by the Schwinger effect in a FRW universe.
Clearly a more precise calculation would be necessary in determining
detailed bounds on primordial electric fields.
It is also important to study what actually happens when the
Schwinger effect becomes relevant; whether the electric fields quickly
decay, or the field decay balances the 
Schwinger production and thus allows the electric field to survive.
Other than from the Schwinger process, the electric field may also be
affected by a gradual decay of the inflaton before it completely
thermalizes the universe, depending on the decay process~\cite{Turner:1987bw}.
We leave a careful exploration of these issues for future work.

\section{Conclusions}
\label{sec:conc}

We showed that primordial electric and magnetic fields do not
necessarily redshift in a radiation-like manner on super-horizon scales.
This is a simple consequence of Maxwell's equations allowing
exchange of power between the two fields.
Given that electric fields stronger than magnetic fields 
are produced in the early universe, the electric fields can render the
magnetic fields to redshift slowly, or even blueshift.
In particular for the standard Maxwell theory, we
showed that the magnetic power scales as $B^2 \propto a^{-6}
H^{-2}$ in the post-inflationary universe until the electric fields
disappear. 

The implication of the induction effect for primordial
magnetogenesis is that the produced magnetic fields continue to be
sourced by the electric fields up until the time of reheating, thus
leading to stronger magnetic strengths than were previously estimated.
The effect is particularly large if the inflation and reheating scales are
well separated, 
and/or the post-inflationary universe has a stiff equation of state,
in which cases the previous estimates are corrected by up to 
37 orders of magnitude.
As an example, we presented a toy model of inflationary magnetogenesis
which produces femto-Gauss magnetic fields on Mpc scales,
combined with a high inflation scale of~$H_{\mathrm{inf}} = 10^{14}\,
\mathrm{GeV}$ and low reheating temperature just above the BBN scale,
with the post-inflation epoch possessing an equation of state~$w = 1/3$. 
This offers a counterexample to the common lore that high-scale
inflation is incompatible with efficient inflationary magnetogenesis;
moreover it opens up the possibility of producing both observable
magnetic fields and gravitational waves from inflation.
It would also be interesting to explore other scenarios of primordial
magnetogenesis by taking into account the correct scaling behavior of
the electromagnetic fields.

We also derived model-independent bounds on primordial magnetic fields
that are supported by the induction effect,
setting a consistency relation between the magnetic field strength and
the reheating scale.
Finally, we briefly commented on the possibility that primordial electric
fields may quench prior to reheating via the Schwinger production of
charged particles,
in which case the magnetic fields would lose support from the electric
fields and thus obey the radiation-like redshifting.
We crudely estimated the condition for the Schwinger process to be
important; a more precise calculation of this effect is an important
task for the future.

Although we have focused on electromagnetic fields throughout this
paper, the induction effect can also be important for addressing the fate
of other gauge fields, such as dark photons, that could have been excited
in the early universe.

\section*{Acknowledgments}

We thank Andr\'e Benevides, Paolo Creminelli, Atish Dabholkar, Ricardo
Z. Ferreira, Daniel G. Figueroa, Daniel Green, Rajeev Kumar Jain,
Mehrdad Mirbabayi, Shinji Mukohyama, Bharat Ratra, and Giovanni
Villadoro for helpful discussions. MSS is supported by Villum Fonden grant 13384. CP3-Origins is partially funded by the Danish National Research Foundation, grant number DNRF90.


\begin{appendix}
\numberwithin{equation}{section}

\setcounter{equation}{0}

\section{Super-horizon matching of classical field}
\label{appA}

Here we match the classical gauge field on super-horizon scales across the inflationary and post-inflationary epochs in the Ratra model, and show that it agrees with the result in Section \ref{sec:power-law}, which was obtained using the more generally applicable method of Bogoliubov transformations. 

Normalizing the mode function as $\tilde{u}_k = I u_k$ (we
drop polarization indices), the equation of motion~(\ref{EoM}) can be written as
\beq
\tilde{u}_k''+\left(k^2 -\frac{I''}{I}	\right)\tilde{u}_k=0~.
\eeq 
The super-horizon regime and the long wavelength regime of  $k^2 \ll \abs{I''/I}$ approximately coincide for reasonable power-law functions $I \propto a^{-s}$, and in this regime the equation has the general solution
\beq
\tilde{u}_k \sim C_1 I +C_2 I\int\frac{d\tau}{I^2}~.
\eeq
In any regime where $I$ is constant (as in the post-inflation regime of the model discussed in Section~\ref{sec:power-law}), the solution will have a constant term and one proportional to $\tau \supset C_3 / (a H)$ (cf.~(\ref{elapsed_ct})), which will be growing in a decelerated expansion phase.

To be more precise, let us consider the model with $I\propto a^{-s}$
during inflation, which by setting the conformal time as $\tau = -1/(a
H_{\mathrm{inf}})$ leads to
\beq
\tilde{u}_k''+\left(k^2 -\frac{s(s-1)}{\tau^2}\right)\tilde{u}_k=0~.
\eeq 
The solution starting from the Bunch-Davies vacuum is given in~(\ref{uk_inf}).
In the super-horizon limit, this can be expanded in terms of $(-k \tau)$
as~\cite{Ferreira:2013sqa} (given that $s$ is not a half-integer),
\beq
u_k=\frac{\tilde{u}_k}{I} = \tilde C_1(k,s)\left\{ 1-
\frac{1}{s+\frac{1}{2}} \left( \frac{-k\tau}{2} \right)^2 + \cdots \right\}
+\tilde D_1(k,s)  \left\{ \left( \frac{-k \tau }{2} \right)^{-2s+1}  +\cdots \right\}
\label{A.4}
\eeq
where the dots indicate higher order terms in the long wavelength approximation and
\bea
\tilde C_1 (k,s) &=& -\frac{i\, \Gamma(-s+\frac{1}{2})}{(2 \pi k)^{1/2}}
 \left( \frac{-k \tau_{\mathrm{end}}}{2} \right)^{s},
\\
\tilde D_1 (k,s) &=& -  \frac{ e^{i s \pi}\, \Gamma(s-\frac{1}{2}) }{(2 \pi k)^{1/2}}  
\left( \frac{- k \tau_{\mathrm{end}}}{2} \right)^s .
\eea
Here the subscript ``end'' denotes the end of inflation, and we have set
$I_{\mathrm{end}} = 1$. 

The mode function after inflation ends and $I$ becomes a constant is  a
sum of plane waves, i.e.~(\ref{uk-Maxwell}), which can be expanded in
terms of $k (\tau - \tau_{\mathrm{end}})$,
\begin{equation}
 u_k =  \frac{1}{(2 k)^{1/2}} 
\left[
\alpha_k (\tau_{\mathrm{end}}) + \beta_k (\tau_{\mathrm{end}})
- i \left\{ \alpha_k (\tau_{\mathrm{end}}) - \beta_k  (\tau_{\mathrm{end}}) \right\} 
k (\tau - \tau_{\mathrm{end}})    
 + \cdots
\right].
\end{equation}
The coefficients $\alpha_k (\tau_{\mathrm{end}})$ and $ \beta_k (\tau_{\mathrm{end}})$
are determined by the matching conditions at~$\tau_{\mathrm{end}}$.
Here, due to considerations pertaining to energy conservation, it is 
$u_k$ and $u_k'$ that has to be matched across the transition. 
Focusing on the case of $s > 1/2$, then the mode function during
inflation~(\ref{A.4}) is dominated by the $\tilde{D}_1$~term and thus
we obtain the coefficients as
\begin{align}
 \alpha_k (\tau_{\mathrm{end}}) + \beta_k (\tau_{\mathrm{end}}) &\simeq
-  \frac{e^{i s \pi }  \, \Gamma (s-\frac{1}{2})  }{\pi^{1/2}}
\left( \frac{- k \tau_{\mathrm{end}}}{2}  \right)^{-s+1},
\\
 \alpha_k (\tau_{\mathrm{end}}) - \beta_k (\tau_{\mathrm{end}}) &\simeq
-  \frac{i\, e^{i s \pi }  \, \Gamma (s+\frac{1}{2})  }{\pi^{1/2}}
\left( \frac{- k \tau_{\mathrm{end}}}{2}  \right)^{-s}.
\end{align}
For the power spectrum of magnetic fields on super-horizon scales after the end of inflation, we then find
\begin{equation}
  \mathcal{P}_B (k) = \frac{k^5}{\pi^2 a^4} |u_k|^2
\simeq \frac{\Gamma (s - \frac{1}{2})^2 k^4}{2 \pi^3 a^4}
\left( \frac{-k \tau_{\mathrm{end}}}{2}  \right)^{-2(s-1)}
\left\{ 1 + (2 s - 1) \frac{\tau - \tau_{\mathrm{end}}}{-\tau_{\mathrm{end}}}  \right\}^2.
\end{equation}
It is easy to verify that this is equivalent to the result in
Eq.~(\ref{EBpower-post-super}) when using $\tau_{\mathrm{end}} =
-1/(a_{\mathrm{end}} H_{\mathrm{inf}})$, and
(\ref{elapsed_ct}) for the elapsed time $(\tau - \tau_{\mathrm{end}})$.

An important observation compared with \cite{Ferreira:2013sqa} is that, when connected to an epoch of $I\propto \abs{\tau}^{\tilde{s}}$ with a different power satisfying $\tilde{s}<-1/2$ in \cite{Ferreira:2013sqa}, the growing solution got matched to the decaying solution, resulting in loss of power at the transition and thus no enhanced magnetic fields. On the other hand in the present case of connecting to an epoch with a constant~$I$, the growing solution gets matched directly on to the growing solution after the transition with no loss of power. 
\end{appendix}

\bibliographystyle{JHEP}
\bibliography{PB}

\end{document}